\documentclass[11pt,a4paper]{article}
\pdfoutput=1
\usepackage{jcappub}
\usepackage[english]{babel}
\usepackage{multirow}
\usepackage{ulem}
\usepackage{fancyhdr}
\usepackage{url}
\usepackage[utf8]{inputenc}
\usepackage{bm}        
\usepackage{amssymb}   
\usepackage{amsmath}
\usepackage{subfig}
\usepackage{caption,booktabs}
\usepackage{hyperref}
\usepackage{xcolor}
\usepackage{lineno}
\usepackage[symbol]{footmisc}
\usepackage{lettrine}
\usepackage{soul}
\newcommand{\de}{\mathrm d}

\newcommand{\g}{$\gamma$}
\newcommand{\bi}{\begin{itemize}}
\newcommand{\ei}{\end{itemize}}
\newcommand{\be}{\begin{equation}}
\newcommand{\ee}{\end{equation}}
\newcommand{\nn}{\mathcal{N}}
\newcommand{\Fermi}{\textit{Fermi}-LAT}

\newcommand{\phys}{physical}
\newcommand{\pheno}{phenomenological}

\usepackage{graphicx}
\usepackage{float}
\usepackage{url}
\usepackage{hyperref}
\usepackage{amsmath}
\usepackage{amssymb,bm}

\newcommand{\rmagic}{\textsc{redMaGiC}}
\def\bea{\begin{eqnarray}}
\def\eea{\end{eqnarray}}
\def\be{\begin{equation}}
\def\ee{\end{equation}}

\usepackage{xcolor}
\usepackage{subcaption}

\title{Multi-Tracer Cross-Correlations of the Unresolved $\gamma-$Ray Sky}

\author[1,2,3]{B.\ Thakore,\footnote[2]{bhashinashish.thakore@unito.it/b.thakore@uva.nl}}
\author[1,2]{M.\ Regis,}
\author[4]{M.\ Negro,}
\author[1,2,5]{S.\ Camera,}
\author[7,8]{D.\ Gruen,}
\author[1,2]{N.\ Fornengo,}
\author[9,10]{and A.\ Roodman}

\affiliation[1]{Dipartimento di Fisica, Universit\`a degli Studi di Torino,\\Via P.\ Giuria 1, 10125 Torino, Italy}
\affiliation[2]{INFN -- Istituto Nazionale di Fisica Nucleare, Sezione di Torino,\\Via P.\ Giuria 1, 10125 Torino, Italy}
\affiliation[3]{GRAPPA (Gravitation Astroparticle Physics Amsterdam), University of Amsterdam,\\Science Park 904, 1098 XH Amsterdam, The Netherlands}
\affiliation[4]{Department of Physics \& Astronomy, Louisiana State University,\\Baton Rouge, LA 70803, USA}
\affiliation[5]{INAF -- Istituto Nazionale di Astrofisica, Osservatorio Astrofisico di Torino,\\Strada Osservatorio 20, 10025 Pino Torinese, Italy}
\affiliation[7]{University Observatory, Faculty of Physics, Ludwig-Maximilians-Universität,\\Scheinerstr.\ 1, 81679 Munich, Germany}
\affiliation[8]{Excellence Cluster ORIGINS,\\Boltzmannstr.\ 2, 85748 Garching, Germany}
\affiliation[9]{SLAC National Accelerator Laboratory,\\Menlo Park, CA 94025, USA}
\affiliation[10]{Kavli Institute for Particle Astrophysics \& Cosmology,\\P.O.\ Box 2450, Stanford University, Stanford, CA 94305, USA}

\abstract{Our understanding of the \g-ray sky has greatly advanced, yet studying the unresolved \g-ray background (UGRB) can unveil the nature of the faintest \g-ray source populations in the Universe. Statistical cross-correlations between the UGRB and tracers of large-scale cosmic structure allow us to infer which sources contribute the most to this emission. In this work, we examine the angular correlation between the UGRB and the matter distribution traced by galaxies, using twelve years of \textit{Fermi} Large Area Telescope (LAT) observations along with three years of Dark Energy Survey (DES) data. We detect a correlation with a signal-to-noise ratio of 7.85, primarily driven by large angular scales. 
We then perform a multi-tracer analysis that combines this measurement with the cross-correlation between \g\ rays and DES weak lensing. The two single-tracer results are mutually consistent, and their combination yields a total significance of 10.31, firmly establishing the extragalactic origin of the UGRB. Intriguingly, the properties inferred for the sources contributing to the UGRB show departures from those of the resolved \g-ray population, suggesting that the faint end of the \g-ray sky is not a simple extrapolation of currently resolved sources.
}
\date{November 2025}

\begin{document}

\maketitle

\section{Introduction}
High-energy astrophysics is an endeavour that can provide insights into the physics of the cosmos, from cosmology to fundamental physics. 
Stemming from extremely violent events in the Universe, \g-rays act as messengers, probing populations of sources and processes across cosmic time and scales.


In the GeV regime, the \g-ray sky is populated by emissions from many astrophysical sources, both Galactic, such as binary and isolated neutron stars, pulsar wind nebulae and supernova remnants, and extragalactic, like active galactic nuclei (AGNs), mostly blazars \citep[see][for the latest \g-ray source catalog by the \textit{Fermi} Large Area Telescope]{abdollahi2022incremental}. Additionally, high-energy cosmic-ray interactions with the Galactic interstellar medium and radiation fields create a complex foreground that can mask fainter contributing background sources \citep{2016ApJS..223...26A}. The expected signals from distinct, faint populations are therefore difficult to isolate against this noisy backdrop. In addition to this, faintest sources that are below the \Fermi\ detection threshold cannot be detected directly. These sources constitute the Unresolved \g-ray Background \citep[UGRB][]{2015ApJ...799...86A, 2016arXiv160104323D}.

A powerful method for separating different contributions to the UGRB is to cross-correlate UGRB maps with other tracers of the large-scale structure of the Universe~\cite{camera2013novel,Fornengo2014}. Useful tracers include the clustering of galaxies and galaxy clusters \cite{ando2014mapping,xia2015tomography,regis2015particle,cuoco2015dark,shirasaki2015cross,cuoco2017tomographic,Ammazzalorso:2018evf,paopiamsap2024constraints,branchini2017cross,hashimoto2019measurement,colavincenzo2020searching,tan2020bounds,Pinetti:2025hgd}, weak lensing~\cite{camera2015tomographic,shirasaki2014cross,troster2017cross,shirasaki2016cosmological,shirasaki2018correlation,DES:2019ucp,Zhang:2026ysp}, as well as the lensing effect of the Cosmic Microwave Background (CMB) \cite{fornengo2015evidence,feng2017planck}. These observables trace the distribution of matter and galaxies across cosmological distances and therefore encode complementary information about the origin of \g-ray emission.

Cross-correlating the UGRB with tracers of large-scale structure such as galaxy clustering and weak lensing is particularly promising because it leverages the three-dimensional information encoded in the combination with galaxy surveys: angular scale, energy, and redshift dependence. Different populations of astrophysical sources tend to have smooth, power-law–like energy spectra with different slopes, and other contributors, such as dark matter (DM), can show characteristic spectral features or cutoffs. 
The cosmological density of different \g-ray populations scales differently with the star formation history or clustering evolution, thus showing different redshift dependence.
The way different large-scale structure probes traces the underlying matter distribution is different, leading to a different angular dependence for their correlation with \g-ray sources.
By jointly modelling the angular, spectral and redshift behaviour of the cross-correlation signal, we can constrain the clustering properties, luminosity functions and redshift distributions of the various contributors to the UGRB, refining our picture of AGNs, distant star-forming galaxies (SFGs), and other yet-to-be-resolved \g-ray emitters~\cite{camera2013novel}. 
A key contributor to the UGRB is the blazar population: blazars are a class of AGNs whose jet's orientation aligns with the line of sight within 5 degrees \cite{ajello2015origin}. Blazars have been shown to dominate almost exclusively the small-scale anisotropy of the UGRB as measured by \Fermi\ \cite{Ackermann:2018wlo, korsmeier2022flat}. Accurately modelling blazars’ clustering and spectra is therefore essential when interpreting cross-correlation measurements.

In this work, we present a study of cross-correlations between the UGRB and galaxy clustering and extend our analysis from Ref. \cite{thakore2025high} using a multi-tracer approach that combines galaxy clustering with weak lensing. The multi-tracer framework increases the sensitivity of our measurements.
By exploiting the complementary redshift kernels and bias dependencies of galaxy clustering and lensing, the multi-tracer method helps break degeneracies such as those between source bias, redshift distribution, and intrinsic UGRB properties, thereby improving constraints on the astrophysical composition of the UGRB. The galaxy clustering observable we consider is the projected galaxy overdensity, measured in tomographic redshift bins to retain redshift sensitivity. Over the years, several observational attempts at measuring the cross-correlation between the UGRB and galaxy clustering have been carried out  
\cite{xia2015tomography,regis2015particle,cuoco2015dark,shirasaki2015cross,cuoco2017tomographic,Ammazzalorso:2018evf,paopiamsap2024constraints}. Here, we build on these efforts using 12 years of \g-ray data from \Fermi\ \cite{Fermi-LAT:2019yla,abdollahi2022incremental} together with three-year (Y3) galaxy clustering measurements from the Dark Energy Survey (DES).

The paper is structured as follows. In Section \ref{sec:data}, we present the survey and telescope data sets used in our analysis. Section \ref{sec:theomod} outlines the theoretical framework for modelling the cross-correlations between \g-ray intensity fluctuations and gravitational tracers of large-scale structure. Section \ref{sec:results_galxgam} reports our measurements of the galaxy–\g-ray cross-correlations and discusses their phenomenological and physical interpretations. Finally, in Section \ref{sec:multitracer_results}, we introduce our multi-tracer analysis, which combines weak lensing and galaxy clustering information to assess the joint constraining power of these probes. Section~\ref{sec:concl} concludes.

\section{Data} \label{sec:data}

\subsection{DES - \rmagic} \label{sec:redma}
The galaxy sample employed in this work is selected using the 
\textsc{redMaGiC} algorithm~\cite{rozo2016redmagic}, applied to the DES Y3 data. The \textsc{redMaGiC} method identifies 
Luminous Red Galaxies by exploiting the well-defined 
magnitude-color-redshift relation of red-sequence galaxies, which is 
calibrated with an overlapping spectroscopic reference sample. These are lens galaxies, which are the foreground galaxies that bend light from distant soure galaxies to create the lensing effect. Galaxies 
are required to satisfy a luminosity threshold and to provide an 
adequate fit to the \textsc{redMaGiC} template relation, quantified by 
a $\chi^2_{\mathrm{RM}}$ statistic, where RM indicates \rmagic galaxies. Only sources with 
$\chi^2_{\mathrm{RM}} < \chi^2_{\max}$ are retained, where 
$\chi^2_{\max}$ is chosen to enforce a nearly constant comoving number 
density. Applying this selection to DES Y3 yields a catalogue containing 
approximately $2.6 \times 10^{6}$ galaxies.  

The observed number density of \rmagic\ galaxies correlates with several survey observing conditions, potentially introducing systematic fluctuations 
into clustering measurements ~\cite{faga2025dark}. To correct for these effects, each galaxy 
is assigned a weight equal to the inverse of the local angular selection 
function, following the methodology validated in~\cite{rodriguez2022dark}. The tomographic redshift distribution for the \rmagic\ lens galaxies is shown at the bottom of Fig. \ref{fig:nz_distros}. The intervals for the five \rmagic\  redshift bins are $0.15 \leq z_1 \leq 0.35$, $0.35 \leq z_2 \leq 50$, $0.50 \leq z_3 \leq 0.6$, $0.65 \leq z_4 \leq 0.8$, and $0.80 \leq z_5 \leq 0.9$.

\subsection{DES - Tangential Shear catalogue}
For the weak lensing measurements we use the 
\textsc{Metacalibration} shear catalogue 
\cite{sheldon2017practical,huff2017metacalibration}, included in the DES 
Y3 shape catalogue~\cite{gatti2021dark}. This catalogue is constructed from a 
subset of objects in the DES \textsc{Gold} sample~\cite{sevilla2021dark}, 
and provides calibrated galaxy ellipticities and tomographic redshift 
bin assignments. The \textsc{Metacalibration} method applies artificial 
shears to images to measure the response of the shear estimator, allowing 
for correction of model and noise biases via a mean response factor. 
It has been validated to the part-per-thousand level in the absence of 
blending and to the part-per-hundred level under DES Y3 conditions. After 
bias corrections, the catalogue contains $\sim 10^{8}$  galaxies.  

\begin{figure*}[!htbp]
\centering
  \includegraphics[width=0.8\textwidth]{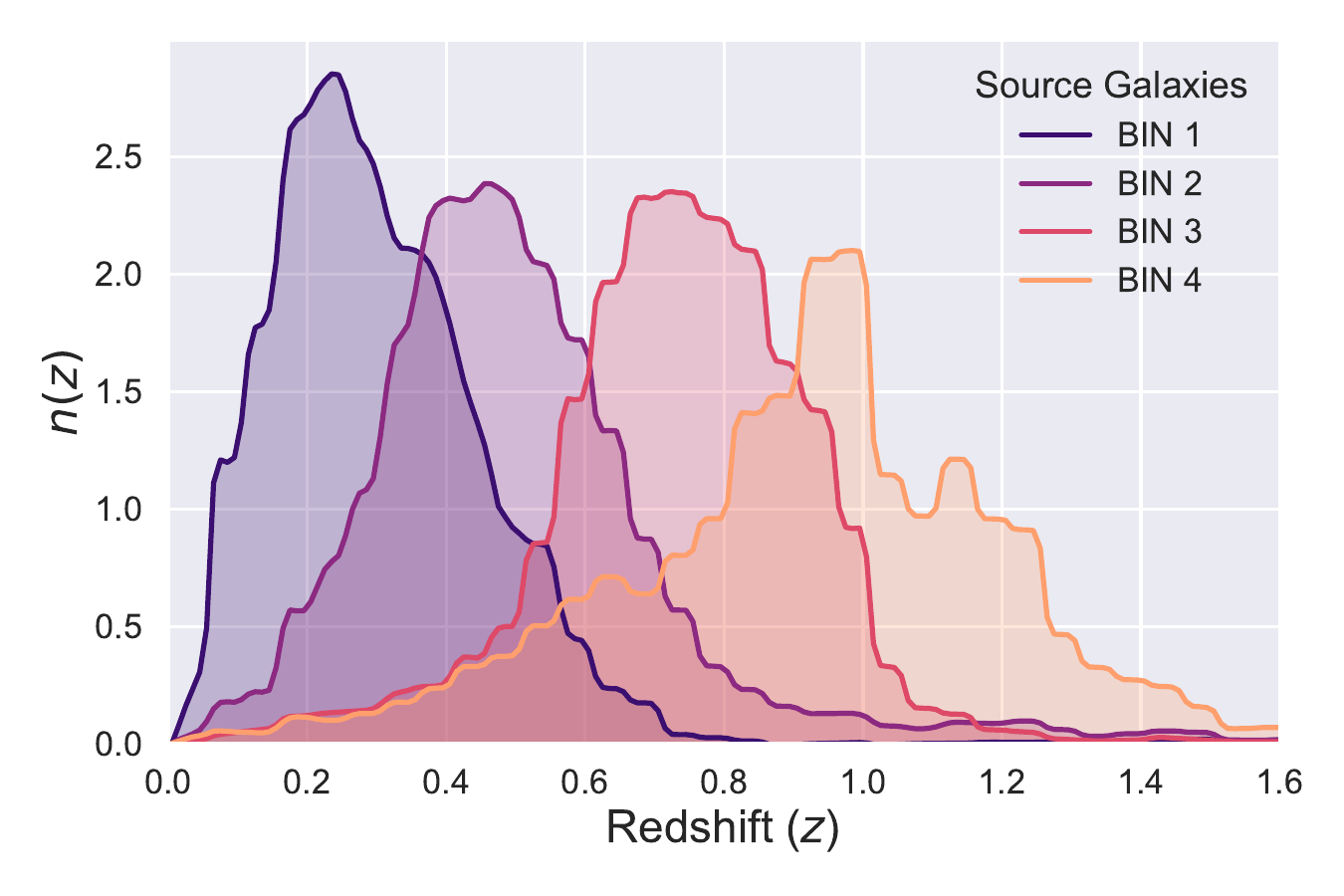}
  \includegraphics[width=0.8\textwidth]{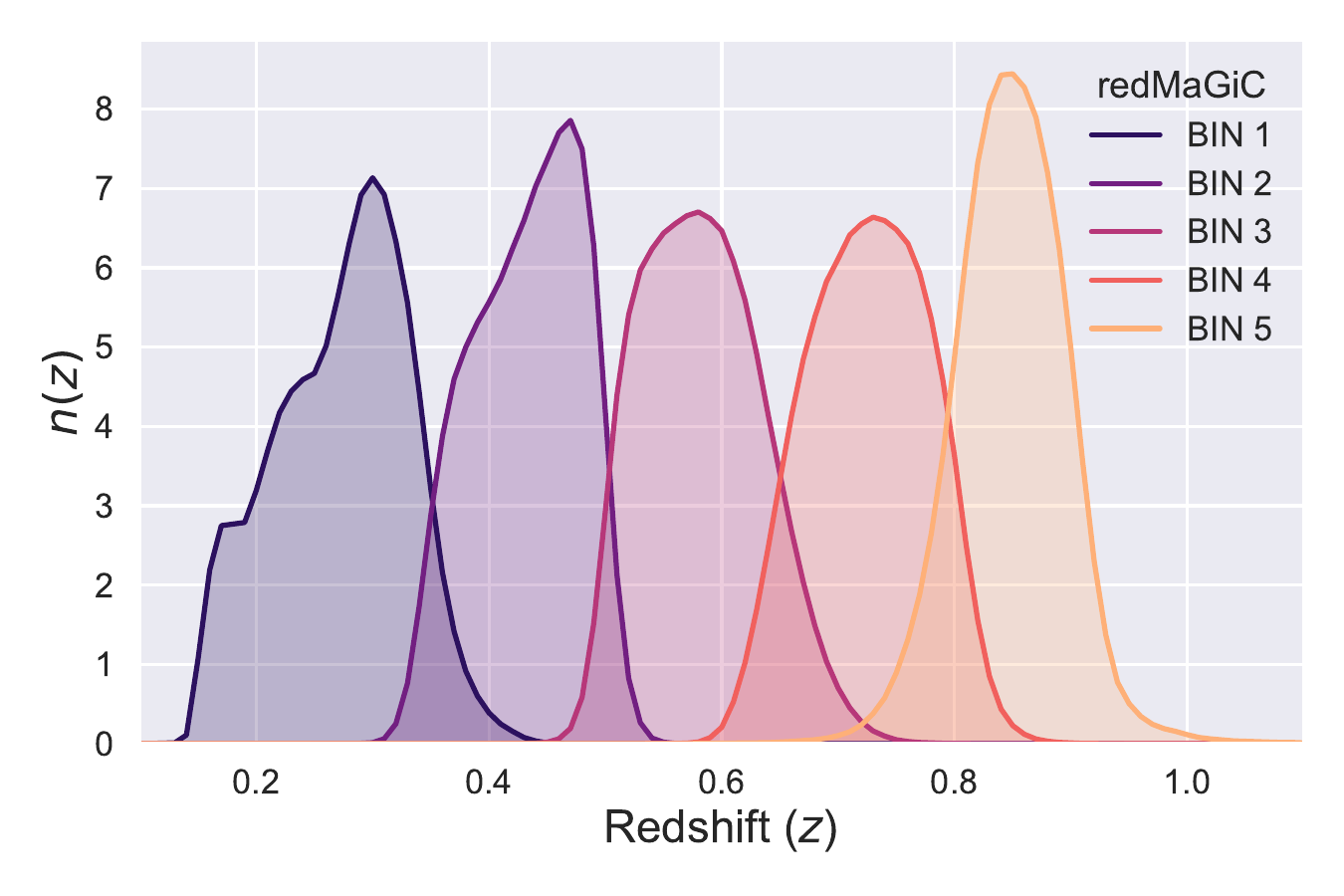}
  \caption{The redshift distributions for the DES Y3 source and 
  \rmagic\ lens galaxies. The data used to generate the plots has been taken from the publicly available DES database (\url{https://des.ncsa.illinois.edu/releases/y3a2/Y3key-catalogues}) and has been normalised such that $\int n(z)dz = 1$ for both samples.} 
  \label{fig:nz_distros}
\end{figure*}

Calibration of source redshifts combines the Self-Organizing Maps method ~\cite{buchs2019phenotypic,myles2021dark} with clustering 
redshifts ~\cite{gatti2022dark}, with uncertainties characterized 
through ensembles of redshift realizations and simulations with 
\textsc{Balrog} sources~\cite{suchyta2016no,everett2022dark}. Shear ratio 
measurements~\cite{sanchez2022dark} provide an additional, largely 
independent, validation of the redshift distributions. The redshift distribution is shown in the top image of Fig.\ \ref{fig:nz_distros}. The four DES Y3 
tomographic bins have mean redshifts 
$\langle z_1 \rangle = 0.339$, 
$\langle z_2 \rangle = 0.528$, 
$\langle z_3 \rangle = 0.752$, and 
$\langle z_4 \rangle = 0.952$~\cite{myles2021dark}.

\subsection{Fermi-LAT}

\begin{figure}
    \centering
    \includegraphics[width=0.85\linewidth]{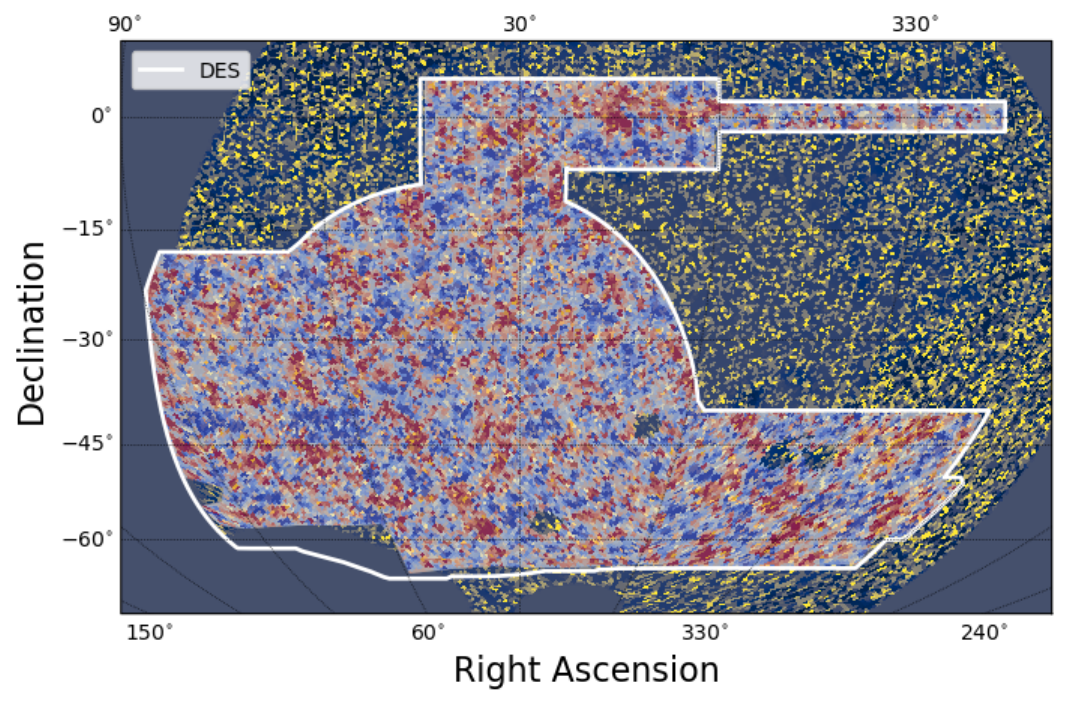}
    \caption{The DES Y3 mass map \cite{jeffrey2021dark} along with the overlapping region in the \Fermi\ 12 year \g-ray background with energies between 9.120 and 17.38 GeV. The plot is in a Mollweide projection with equatorial coordinates, and the \g-ray map has been downgraded to $\mathrm{N_{side}}=256$ for visualization purposes. }
    \label{fig:maps}
\end{figure}

\Fermi\ \citep{2009ApJ...697.1071A} provides over a decade of all-sky continuously monitoring in the energy band between about 50 MeV and above 300 GeV. In this work, the data reduction and the extraction of the UGRB component are performed as in Ref.\ \cite{thakore2025high}, which measured the cross-correlation between the UGRB and tangential shear, and to which we defer for the details. Here we summarize the main points of the data analysis. 
We use 12 years of \Fermi\ Pass 8 (R3) \citep{2013arXiv1303.3514A, 2018arXiv181011394B}, spanning 
August 2008 to August 2020. With the \texttt{Fermitools} (v2.2.0)\footnote{\url{https://fermi.gsfc.nasa.gov/ssc/data/analysis/documentation/}}, we select 
\textsc{sourceveto\_v2} events of types PSF1+2+3, excluding the PSF0 
subset to improve angular resolution. This event class provides an 
optimal balance between acceptance and residual contamination, yielding a 
clean photon dataset well-suited for cross-correlation studies. The maps correspond to nine energy bins between $631\,\mathrm{MeV}$ and $1\,\mathrm{TeV}$ (see Tab.\ \ref{tab:enbins}) and are generated in \texttt{HEALPix} format with $N_{\rm side}=1024$, matching the 
angular resolution of the DES Y3 weak lensing data.\footnote{The maps are publicly available and can be found in the github repository here: \url{https://github.com/BhashinT/Fermi-LAT-Gamma-Ray-Maps-12-years-}}.
Following the procedure adopted in several previous works \citep{Ackermann:2018wlo, thakore2025high}, we extract the UGRB emission in each energy bin by applying a latitude mask (excluding $|b|<30^\circ$), and masking detected point sources from the 4FGL-DR3 catalogue~\cite{abdollahi2022incremental} with energy-dependent masks accounting for source flux and the PSF (see details in Ref.\ \citep{thakore2025high}). After fitting and subtracting the diffuse Galactic foreground template~\cite{2018PhRvL.121x1101A} outside the mask, the resulting masked flux maps (in unit of ph/cm$^2$/s/sr) are the input of our cross-correlation analysis. An example of the cleaned $\gamma$-ray intensity map is shown in Fig.\ \ref{fig:maps}, along with the DES Y3 footprint. 
\begin{table*}[t]
\centering
\resizebox{\textwidth}{!}{%
\begin{tabular}{llllllllll}
\hline
 & \multicolumn{9}{c}{Bin number} \\
\cline{2-10}
 & 1 & 2 & 3 & 4 & 5 & 6 & 7 & 8 & 9 \\
\hline\hline
$E_{\rm min}$ [GeV] & 0.631 & 1.202 & 2.290 & 4.786 & 9.120 & 17.38 & 36.31 & 69.18 & 131.8 \\
$E_{\rm max}$ [GeV] & 1.202 & 2.290 & 4.786 & 9.120 & 17.38 & 36.31 & 69.18 & 131.8 & 1000 \\
$\theta_{\rm cont}$ 68\% [deg] & 1.00 & 0.58 & 0.36 & 0.22 & 0.15 & 0.12 & 0.11 & 0.10 & 0.10 \\
Photon counts & 142{,}208 & 592{,}345 & 491{,}506 & 212{,}579 & 87{,}486 & 37{,}551 & 11{,}797 & 3{,}676& 1{,}591 \\
$\langle I_a \rangle$ [$10^{-7}\,\mathrm{cm^{-2}\,s^{-1}\,sr^{-1}}$] & 5.59 & 2.13 & 0.978 & 0.374 & 0.154 & 0.0672 & 0.0197 & 0.0059 & 0.0032 \\
\hline
\end{tabular}
}
\caption{Gamma-ray energy bins used in the analysis, showing the 68\% containment angles $\theta_{\rm cont}$ of the {\it Fermi}-LAT PSF, the photon counts in the unmasked {\it Fermi} area for each bin, and the corresponding average measured intensity.}
\label{tab:enbins}
\end{table*}

\hspace{1em}


\section{Theoretical models \label{sec:theomod}}
The first part of this study concerns the 2-point angular correlation function between fluctuations in galaxy number counts and the UGRB, while the second part combines this cross-correlation with that of the UGRB with the tangential shear~\cite{thakore2025high}. In both cases, we want to test and build upon the notion that the \g-ray fluctuation field in the Universe traces the cosmological distribution of matter.
For the theoretical model of the correlation between the \Fermi\ UGRB and DES tangential shear, we refer the reader to Ref.\ \cite{thakore2025high}. Here we discuss the theoretical model of the correlation between the \Fermi\ UGRB and the galaxy distribution.

The cross-correlation in physical (i.e.\ configuration) space of galaxies in redshift bin $r$ with the UGRB in energy bin $a$ can be obtained as a transformation from the harmonic space cross-correlation power spectrum $C_\ell^{ar}$ as 
\be
\hat\Xi^{ar}(\theta) = \sum_\ell \frac{2\ell+1}{4\pi} C_\ell^{ar}\, W_\ell^a\,P_\ell(\cos\theta)\;,
\label{eq:Cl2xi}
\ee
where $\theta$ is the angular separation in the sky, $P_\ell(\cos\theta)$ is the Legendre polynomial, and $W_\ell^a$ is the beam window function computed from the Legendre transform of the \Fermi\ PSF (c.f. Appendix II in Ref. \cite{Ackermann:2018wlo}). The beam window function accounts for the finite resolution of the detector. We neglect DES and pixel smoothing here since they operate on scales much smaller than the \Fermi\ PSF. 

In both parts of this work, we base our theoretical formulations on the halo model approach~\cite{asgari2023halo}. Such a framework assumes that the matter content of the Universe structure resides in virialised dark matter halos, and can be distinctly split into two different components: the 1-halo (1h) term and the 2-halo (2h) term. The 1h term captures cross-correlations between two points of origin within the same halo (thereby dominating at smaller angular scales), while the 2h term concerns cross-correlations between two different halos and provides information about the large-scale matter distribution of the Universe. 
We broadly employ two kinds of theoretical approach, a phenomenological model and a physical model, described in more detail in the subsections below. 

\subsection{Phenomenological model for the galaxy-UGRB cross-spectrum}
We define a phenomenological model as a generic description in terms of a cross-correlation function that does not explicitly attempt to separate the signal into its individual physical components, but captures its main characteristics. Accordingly, we define our first phenomenological model as a Power-Law (PL) in energy and redshift as
\begin{equation}
\begin{split}
\Xi_{\rm PL}^{ar}(\theta) \  = &\Big[A_1  (E_a/E_0)^{-\alpha_1} \Big(\frac{1+z_r}{1+z_0}\Big)^{\beta_1}\ \hat{\Xi}_\textrm{PSF-like}^{a}(\theta) \\
&+A_2 (E_a/E_0)^{-\alpha_2} \Big(\frac{1+z_r}{1+z_0}\Big)^{\beta_2}\ \hat\Xi_\textrm{2h-like}^{ar}(\theta)\Big]\frac{\Delta E_a}{\langle I_a \rangle}\;,
\label{eq:phenopl}
\end{split}
\end{equation}
where $E_a$ and $z_r$ denote the central values (defined as the geometric mean of the corresponding bin extrema) of the energy (measured in  GeV) and redshift bins, $E_0 = 13.7\, \mathrm{GeV}$ is a pivot energy chosen as the geometric mean of the energy bin centres,  $\Delta E_a$ is the width of the $a$-th energy bin and $\langle I_a \rangle$ is the corresponding measured photon flux, provided in Table~\ref{tab:enbins}. Similarly to $E_0$, we have also defined a pivot redshift as a geometric mean of the centres of the four redshift bins at $z_0 = 0.507$.\footnote{The pivot energy and redshift values are arbitrary, even though the choice we made helps the numerical stability.}

$\hat{\Xi}_\textrm{PSF-like}^{a}(\theta)$ denotes the Legendre transform of the \Fermi\ PSF 
(i.e., assuming $C_\ell^{ar} = 1$ in Eq.\ (\ref{eq:Cl2xi})), as appropriate for the 1-halo term of point-like objects. 
The $C_\ell^{ar}$ providing $\hat{\Xi}_\textrm{2h-like}^{ar}(\theta)$ turns out to have very similar angular shapes in different physical models and in different energy or redshift  bins (while on the opposite it might have dramatically different amplitudes in different bins or different models). This is because the angular behaviour is dictated by the linear matter power spectrum, i.e., because all matter tracers show a similar large-scale pattern. For the sake of concreteness, we pick one of the angular power spectra of the physical model discussed below, checking that different choices have a negligible impact on our results.
Note that the term in square brackets in Eq.~\ref{eq:phenopl} is differential in energy. 
Correlation functions marked with a hat carry flux units, whereas those without a hat are normalised to the \g-ray\ flux and are therefore dimensionless. 

The normalisation parameters ($A_1$ and $A_2$), the spectral indices ($\alpha_1$ and $\alpha_2$), 
and the redshift evolution indices ($\beta_1$ and $\beta_2$) are treated as free parameters in the model. 
In this phenomenological framework, the relative amplitude between the PSF-like and large-scale components is therefore free (i.e., given by the ratio $A_1/A_2$) and the ensuing combination of the 1h and 2h terms sets the angular dependence of the signal. Astrophysical \g-ray sources generally exhibit energy spectra that can be well described by a power law, the assumption adopted in Eq.\ (\ref{eq:phenopl}). For simplicity, we assume a PL scaling also for the redshift dependence.

Since several sources in the \Fermi\ 4FGL catalogue are also described using a Log-Parabola (LP) shape in energy~\cite{Fermi-LAT:2019yla}, we include, in addition to the PL phenomenological model, also an LP model as
\begin{equation}
\begin{split}
\Xi_{\rm \mathrm{LP}}^{ar}(\theta) \  = &\Big[A_1  (E_a/E_0)^{-\alpha_1-\gamma_1\log_{10} (E/E_0)} \Big(\frac{1+z_r}{1+z_0}\Big)^{\beta_1}\ \hat{\Xi}_\textrm{PSF-like}^{a}(\theta) \\
&+A_2 (E_a/E_0)^{-\alpha_2-\gamma_2\log_{10}(E/E_0)} \Big(\frac{1+z_r}{1+z_0}\Big)^{\beta_2}\ \hat\Xi_\textrm{2h-like}^{ar}(\theta)\Big]\frac{\Delta E_a}{\langle I_a \rangle}\;,
\label{eq:phenolp}
\end{split}
\end{equation}
where $\gamma_1$ and $\gamma_2$ denote the spectral indices characterising the deviation from a power-law behaviour, hereafter referred to as curvature indices.

\subsection{Physical model for the galaxy-\g-ray correlation}
A physical model for the angular power spectrum of the cross-correlation between \g-ray\ sources and the galaxies can be expressed as~\cite{Fornengo2014}
\be
C_\ell^{ar} = \int \de E\, \de z\, \frac{1}{H(z)} 
\frac{W_\textrm{\g}^a(E,z)\, W_{\rm galaxies}^r(z)}{\chi^2(z)} 
P_{\gamma g}\!\left[k=\frac{\ell}{\chi(z)}, z\right]\;,
\label{eq:clgen}
\ee
where $\chi(z)$ is the comoving distance at redshift $z$, satisfying $\de z / \de \chi = H(z)$ in a flat Universe, with $H(z)$ being the Hubble parameter. 
$W_\textrm{\rm gamma}^a(E,z)$ and $W_{\rm galaxies}^r(z)$ are the window functions describing the redshift distributions of \g-rays and galaxies, respectively, while $P_{\gamma g}$ is the three-dimensional cross-power spectrum between a given \g-ray-emitting population contributing to the UGRB and the DES galaxy distribution. 
Within the Limber approximation, the physical wavenumber $k$ and the angular multipole $\ell$ are related by $k = \ell / \chi(z)$. 
Cosmological parameters entering Eq.\ (\ref{eq:clgen}) are from Ref.\ \cite{DES:2021wwk}.


The window function for \rmagic\ galaxies is given by $W^r(z)=H(z)/c\,\de n(z)/\de z$, with the source redshift distribution described in Sec.\ \ref{sec:redma}.

We assume blazars (BLZ) to dominate the \g-ray\ source population. 
The \g-ray\ window function depends on the \g-ray\ luminosity function (GLF). 
We decompose the GLF,
$\Phi_S(L_{\gamma}, z, \Gamma) = \de N_S / (\de L_\gamma\, \de V\, \de \Gamma)$ -- defined as the number of sources per unit luminosity $L_{\gamma}$, comoving volume $V$ at redshift $z$, and photon spectral index $\Gamma$ -- 
into its local form at $z=0$ and an evolutionary factor $e(L_{\gamma}, z)$, namely
\begin{equation}
\Phi_S(L_\gamma, z, \Gamma) = \Phi_S(L_\gamma, 0, \Gamma) \times e(L_{\gamma}, z)\;,
\label{eq:GLF}
\end{equation}
where $L_\gamma$ is the rest-frame luminosity in the $(0.1 - 100)$~GeV range. 
At $z=0$, the GLF reads (as described in Ref. \cite{korsmeier2022flat})
\begin{eqnarray}
\Phi_S(L_\gamma, 0, \Gamma) &=& 
\frac{A}{\ln(10)\, L_\gamma} 
\left[
  \left( \frac{L_\gamma}{L_0} \right)^{\kappa_1} + 
  \left( \frac{L_\gamma}{L_0} \right)^{\kappa_2}
\right]^{-1} 
\exp\!\left[-\frac{(\Gamma - \mu_{\rm BLZ})^2}{2\sigma^2}\right]\;,
\label{eq:phies}
\end{eqnarray}
where $A$ is a normalisation constant, $\kappa_1$ and $\kappa_2$ describe the luminosity dependence of the GLF, 
and the Gaussian term accounts for the distribution of spectral indices $\Gamma$ around a mean value $\mu_{\rm BLZ}$ with dispersion $\sigma$.

The redshift evolution is parametrised as
\begin{equation}
e(L_\gamma, z) =
\left[
  \left( \frac{1+z}{1+z_c} \right)^{-p_1} +
  \left( \frac{1+z}{1+z_c} \right)^{-p_2}
\right]^{-1}\;,
\label{eq:phiz}
\end{equation}
where $z_c$ and $\mu_{\rm BLZ}$ depend on $L_\gamma$~\cite{korsmeier2022flat}. The \g-ray\ window function can then be written as~\cite{DES:2019ucp}:
\be
W_\textrm{{\rm gamma}}^a(E,z) = \chi^2(z)
\int \de \Gamma \int_{L_{\gamma}^{\rm min}}^{L_{\gamma}^{\rm max}} \de L_{\gamma}\,
\Phi_{\rm S}(L_{\gamma}, z, \Gamma)
\frac{\de N}{\de E}\,
e^{-\tau[E(1+z), z]},
\label{eq:win_astro}
\ee
where $\de N / \de E$ is the \g-ray\ photon spectrum (assumed to follow a power law $E^{-\Gamma}$), $\tau$ is the optical depth due to \g-ray\ absorption~\cite{Finke2010}, 
and $L_{\gamma}^{\rm max}$ ensures that only unresolved sources are included. 
Throughout this work, the term ``\g-ray\ spectrum'' refers to the differential photon spectrum, i.e.\ the quantity $\de N / \de E$ in Eq.\ (\ref{eq:win_astro}).

Concerning the structure clustering, we adopt the halo model description. In order to compute the angular cross-correlation between the UGRB and the number of \rmagic\ galaxies, we thus need to describe how galaxies and blazars populate halos. To this aim, we employ the halo occupation distribution (HOD) formalism for galaxies and we will model the average BLZ luminosity of halos.
The HOD approach we follow is described in Ref.\ \cite{Zheng:2004id}, where the HOD of central and satellite galaxies are described separately, since they typically have different formation histories. The expectation value for the number of galaxies is parametrized as
\bea
\langle N_{\rm cen}(M)\rangle &=& \frac{ f_{\rm cen}}{2}\left[1+{\rm erf}\left(\frac{\log M-\log M_{\rm min}}{\sigma_{\rm logM}}\right)\right]\;,\label{eq:HOD1}\\
\langle N_{\rm sat}(M)\rangle &=& \langle N_{\rm cen}(M)\rangle \left( \frac{M}{M_1} \right)^\alpha\;,\label{eq:HOD2}
\eea
with the total number of galaxies being $N_g=N_{\rm cen}+N_{\rm sat}$. 
Five parameters characterise the above two functions:
$M_{\rm min}$ denotes the approximate minimum halo mass required to populate the halo with a \rmagic\ galaxy, with the transition from 0 to 1 central galaxy set by the width $\sigma_{\rm LogM}$; $f_{\rm cen}$ is the fraction of occupied halos (given the selection process of the \rmagic\ algorithm); the satellite occupation is described by a power law (at high masses) with index $\alpha$ and normalization set by $M_1$.
We define a different HOD for each of the five redshift bins described in Sec.\ \ref{sec:redma} and this requires $5\times5$ parameters. We use the best-fit values reported in Ref.\ \cite{Zacharegkas:2022} for the first four redshift bins and assume the same values of the fourth bin also to model the fifth one. Because of this last approximation and because the halo mass function and halo bias used in the following are not identical to the ones in Ref.\ \cite{Zacharegkas:2022}, we allow a rescaling (that turns out to be very moderate $\sim 10-30$\%) for the parameter $f_{\rm cen}$ of each $z$-bin in order to match the effective bias reported in Ref.\ \cite{DES:2021qnp}.

Eqs.\ (\ref{eq:HOD1}) and (\ref{eq:HOD2}) provide the number of galaxies in a halo of mass $M$. The spatial profile of a central galaxy is taken to be a point-source located at the center of the halo, while satellite galaxies are collectively described as following the host-halo profile:
\be
g_g(\bm x|M)=\langle N_{\rm cen}(M)\rangle\,\delta^3(\bm x)+\langle N_{\rm sat}(M)\rangle\,\rho_h(\bm x |M)/M\;,
\ee
where $\int \de^3\bm x\,g_g(\bm x)=\langle N_g(M)\rangle$.
The average number density of galaxies from a given bin $r$ at a given redshift $z$ is given by $\bar n_{g_r}(z)=\int dM\,dn/dM\, \langle N_{g_r}\rangle$.
For the halo mass function $dn/dM$ we use the model in Ref.\ \cite{Sheth:1999mn}, and the halo density $\rho_h$ is taken to follow the NFW profile~\cite{Navarro:1996gj}.

Since the BLZ intrinsic angular sizes are much smaller than the \textit{Fermi}-LAT PSF, and the typical size of the host halos rarely exceeds the PSF, we can assume \g-ray blazars to be point-like sources located at the center of halos.
Their 3D power spectrum of cross-correlation with \rmagic\ galaxies can be thus written as~\cite{cuoco2015dark}:
\bea
 P_{g_r,\gamma_S}^{1h}(k,z) &=& \int_{L_\gamma^{\rm min}}^{L_\gamma^{\rm max}} \de L_\gamma\,\frac{\Phi_S(L_\gamma,z)}{\langle f_{S} \rangle}\,\frac{\de F}{\de E} (L_\gamma,z)\,\frac{\langle N_{g_r}\!(L_\gamma)\,\rangle}{\bar n_{g_r}}\tilde v_g(k|M(L_\gamma))\;, \label{eq:PSBd1} \\
 P_{g_r,\gamma_S}^{2h}(k,z) &=& \left[\int_{L_\gamma^{\rm min}}^{L_\gamma^{\rm max}} \de L_\gamma\,b_{h}(M(L_\gamma))\,\frac{\Phi_S(L_\gamma,z)}{\langle f_{S} \rangle}\,\frac{\de F}{\de E} (L_\gamma,z)\right]\nonumber\\
 &\times &\left[\int_{M_{\rm min}}^{M_{\rm max}} \de M\,\frac{dn}{dM} b_h(M)\,\frac{\langle N_{g_r}\,\rangle}{\bar n_{g_r}} \tilde v_g(k|M) \right] \,P^{\rm lin}(k)\;,\label{eq:PSBd2}
\eea
where $b_{h}$ is the halo bias~\cite{Sheth:1999mn,asgari2023halo}, $F$ is the BLZ \g-ray flux, and $\langle f_{S} \rangle=\int \de L_\gamma\,\Phi_S\, \de F/\de E$.
The product $\langle N_{g_r}\rangle\,\tilde v_g(k|m)$ is the Fourier transform of $g_g(\bm x|M)$. Note that $\langle N_{g_r}\rangle\,\tilde v_g(k=0|M)=\langle N_{g_j}\rangle$. The term within square brackets in the second line of Eqs.\ (\ref{eq:PSBd2}) provides the effective galaxy bias and our model is devised to reproduce the results in Ref.\ \cite{DES:2021qnp}, as mentioned above.
Both Eqs.\ (\ref{eq:PSBd1}) and (\ref{eq:PSBd2}) require the specification of the relation $M(L_\gamma)$ between the mass of the host halo and the \g-ray luminosity of the hosted BLZ, for which we follow Ref.\ \cite{thakore2025high}.
In order to effectively account for the uncertainty in this relation, we can separate the 1h and 2h contributions to the total power spectrum, introducing two different normalisations, and considering the following {\it physical} model,
\begin{align}
\Xi_{\rm phys}^{ar}(\theta)\, \langle I_a \rangle &= 
A_{\rm BLZ}^{\rm 1h}\, \hat\Xi_{\rm BLZ,1h}^{ar}(\theta, \mu_{\rm BLZ}, p_1)
+ A_{\rm BLZ}^{\rm 2h}\, \hat\Xi_{\rm BLZ,2h}^{ar}(\theta, \mu_{\rm BLZ}, p_1)\;.
\label{eq:physmdl}
\end{align}

The parameters constrained via our Markov Chain Monte Carlo (MCMC) analysis are: 
two free normalisations for the 1h and 2h components ($A_{\rm BLZ}^{\rm 1h}$ and $A_{\rm BLZ}^{\rm 2h}$), 
which effectively describe the uncertainty in the overall amplitude $A$ of the GLF in Eq.\ (\ref{eq:phies}) and the normalisation of the $M(L_\gamma)$ relation, as we just mentioned (see also Ref.\ \cite{thakore2025high}); 
the spectral index $\mu_{\rm BLZ}$ in Eq.\ (\ref{eq:phies}), governing the energy dependence of the BLZ \g-ray spectrum; 
and the redshift parameter $p_1$ in Eq.\ (\ref{eq:phiz}).\footnote{We also tested models with independent energy and redshift parameters for the 1h and 2h terms, but found no significant improvement in the fit.}
All remaining GLF parameters are fixed to the best-fit values for BL~Lacs, 
derived from the analysis of \g-ray\ number counts and angular auto-correlation in Ref.\ \cite{korsmeier2022flat} (specifically, the BLL 4FGL+$C_{\rm P}$ fit in their Table~2).


\section{Results and Analysis \label{sec:results_galxgam}}
To measure the cross-correlation between the UGRB and the galaxy distribution, we use \texttt{TreeCorr}'s \textsc{NNCorr} algorithm (see eg. \cite{jarvis2004skewness,jarvis2015treecorr}), which uses the Landy-Szalay \cite{landy1993bias} estimator to calculate the cross-correlation between the DES \rmagic\ catalogues and \Fermi\ images, as well as the random samples associated with both sets of data \footnote{The random catalogues for the galaxies have been obtained from the publicly available DES database found here: \url{https://des.ncsa.illinois.edu/releases/y3a2/Y3redmagic}. The \Fermi\ random images have been generated as explained in Ref.\ \cite{thakore2025high}.}.  The estimator can be written as
\begin{equation}
\label{LSestimator_GalxGam}
    \Xi^{ar}(\theta_h) =  \frac{ \sum_{i,j} \left(D^{r}_{g,i}D^{a}_{\gamma,j} - D^{r}_{g,i}R^{a}_{\gamma,j} - R^{r}_{g,i}D^{a}_{\gamma,j} - R^{r}_{g,i}R^{a}_{\gamma,j}\right)}{\sum_{i,j} R^{r}_{g,i}R^{a}_{\gamma,j}}\;,
\end{equation}
where $D^{r}_{\delta_g,ij}D^{a}_{\gamma,j}$ (`\(D\)' standing for `data') indicate the correlation function in configuration space between the positions of galaxies in pixel $i$ (in redshift bin $r$) relative to a pixel $j$ having an intensity $I^a_j$ (for each energy bin $a$) in the \g-ray maps. The remaining terms indicate cross-correlations used for random (`\(R\)') point subtraction, e.g., $D^{r}_{\delta_g,ij}R^{a}_{\gamma,j}$ denotes the measurement of the correlation of galaxies around random points in the UGRB.
We measure the cross-correlation signal in 12 logarithmically-spaced angular bins spanning radii from $5$ to $600 \,\mathrm{arcmin}$. In Eq.\ (\ref{LSestimator_GalxGam}), $\theta_h$ refers to the geometric center of each angular bin, and the sum is extended only to those pixels whose angular position is inside the correspoding angular bin.
The analysis is carried out in 9 photon energy bins (listed in Table~\ref{tab:enbins}) and the 5 redshift intervals described in Sect.\ \ref{sec:redma}, resulting in a total of 540 measurement bins.
The covariance matrix associated to the measurement is computed as described in Appendix~\ref{sec:cov}.

\begin{table}
\centering
\begin{tabular}{lc|ll|ll|ll|l}
\hline
 & \multicolumn{8}{c}{Subsamples Considered} \\
 \cline{2-8}
 & Full & Low-$z$ & High-$z$ & Low-$E$ & High-$E$ & Small-$\theta$ & Large-$\theta$ & Model\\
\hline
\hline
$\Delta \chi^2_{\rm lp}$ & $61.69$ & $41.31$& $21.87$& $17.66$& $44.64$& 18.13& $54.72$& LP\\
${\rm SNR}_{\rm lp}$ & $7.85$ & $6.42$ & $4.68$ & $4.20$ & $6.68$ & $4.39$ & $7.42$ & LP \\
\hline
$\Delta \chi^2_{\rm pl}$ & $33.63$& $17.34$& $17.62$& $10.14$& $23.77$& $10.14$& $33.98$& PL \\
${\rm SNR}_{\rm pl}$ & $5.70$& $4.10$& $4.22$& $3.33$ & $4.95$& $3.08$& $5.76$& PL  \\
 \hline
$\Delta \chi^2_{\rm phys-BLZ}$ & $34.84$ & $15.72$& $19.76$& $12.97$& $34.59$& $12.97$& $34.59$& Physical\\
${\rm SNR}_{\rm phys-BLZ}$ & $5.89$ & $4.00$& $4.28$& $3.00$ & $5.28$ & $3.60$& $5.89$& Physical\\
 \hline
\end{tabular}
\caption{$\Delta \chi^2_{\rm mod}$ and ${\rm SNR}_{\rm mod}$ computed for the \pheno\ and \phys\ models, using either the full data set or the various subsamples discussed in the text. For the Low-$z$ case we selected the first two redshift bins, while for the High-$z$ case the last three bins; the Low-$E$ subset refers to the first four energy bins, i.e., energies below 9 GeV, while the High-$E$ to the bins at higher energies; finally, the {Small-$\theta$/Large-$\theta$} cases correspond to data points below/above the $68\%$ containment angle of the \Fermi\ PSF.}  
\label{tab:chi2comp}
\end{table}

\begin{figure*}[htbp!]
\centering
    \includegraphics[width=0.45\textwidth]{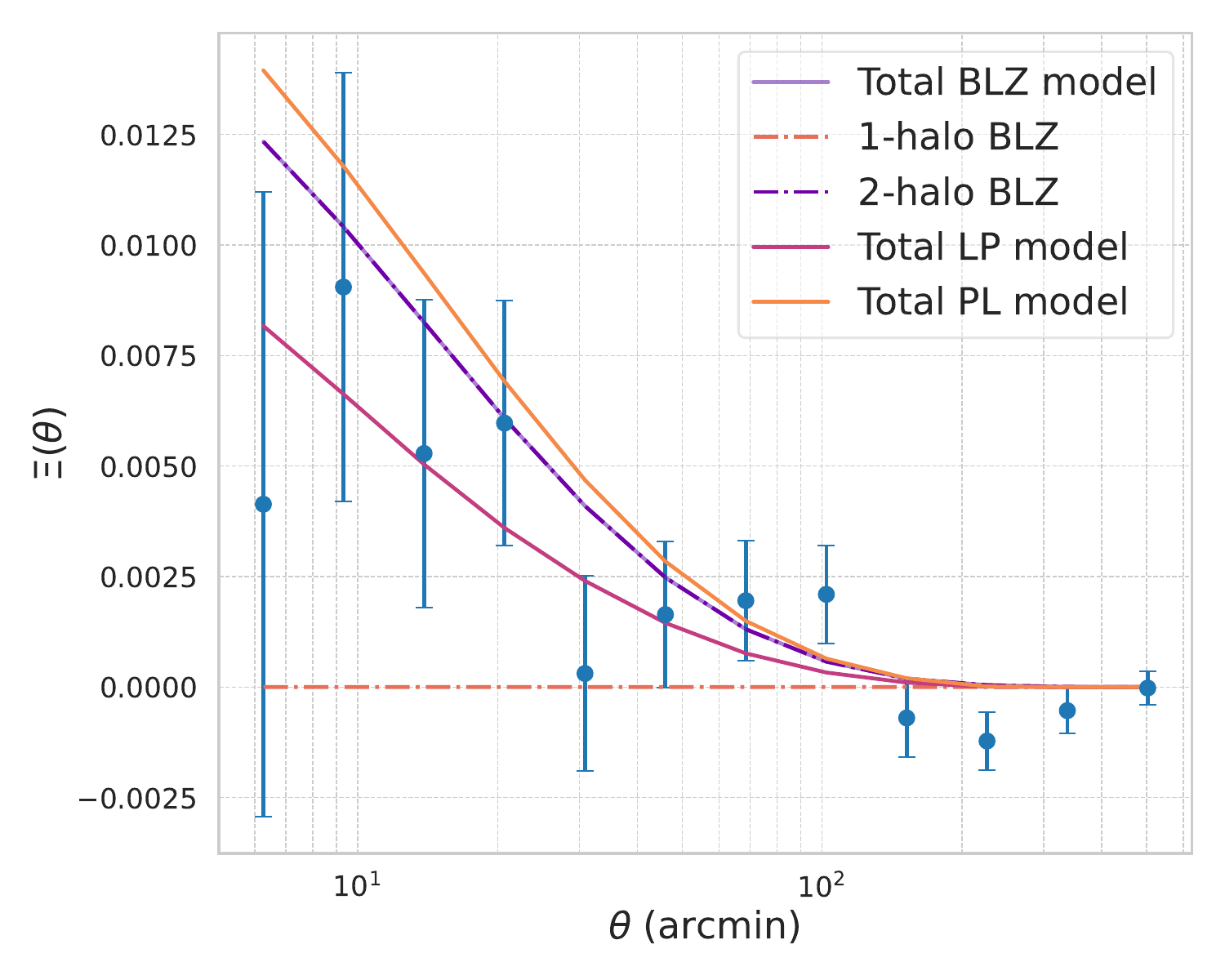}
    \includegraphics[width=0.45\textwidth]{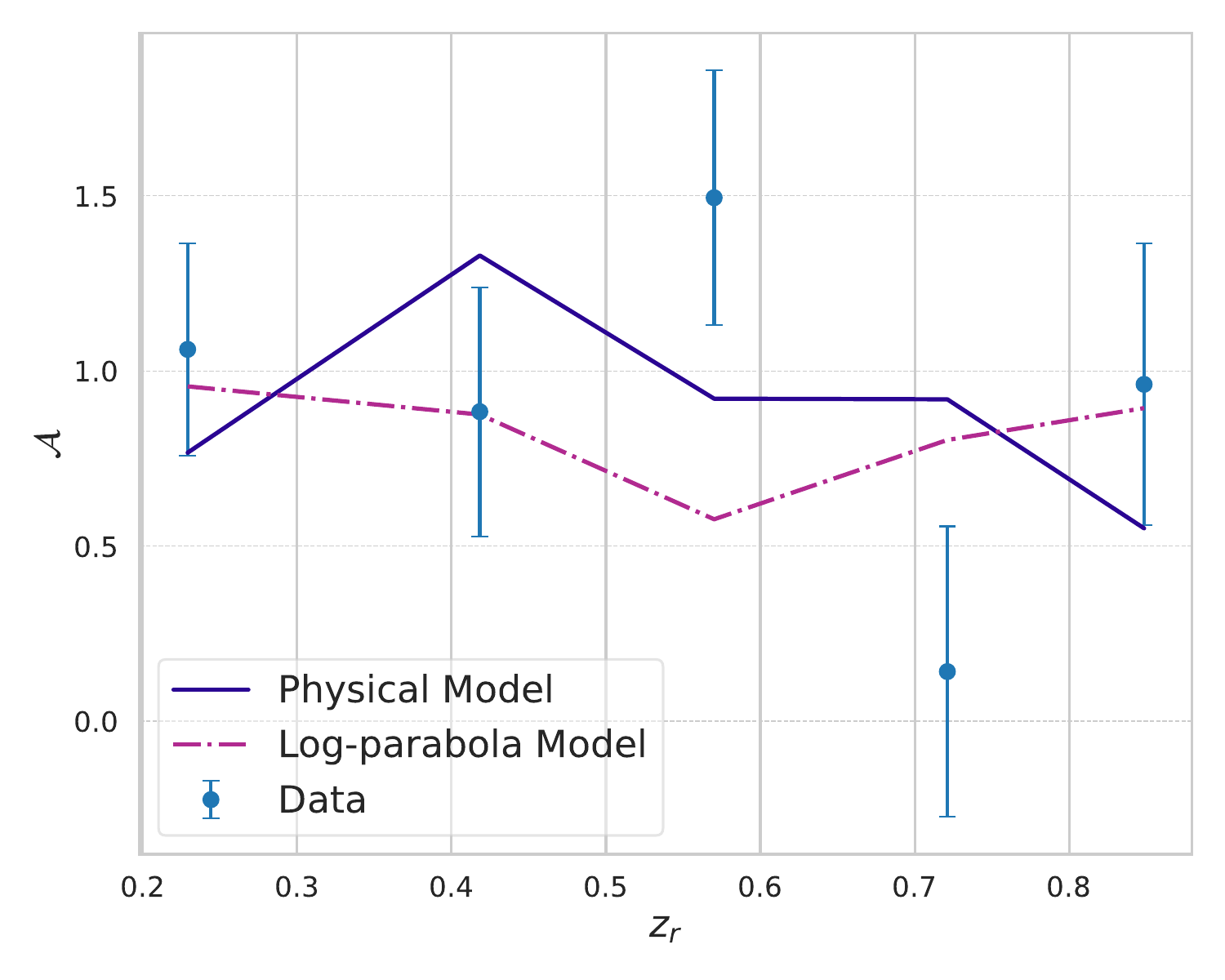}
    \includegraphics[width=0.45\textwidth]{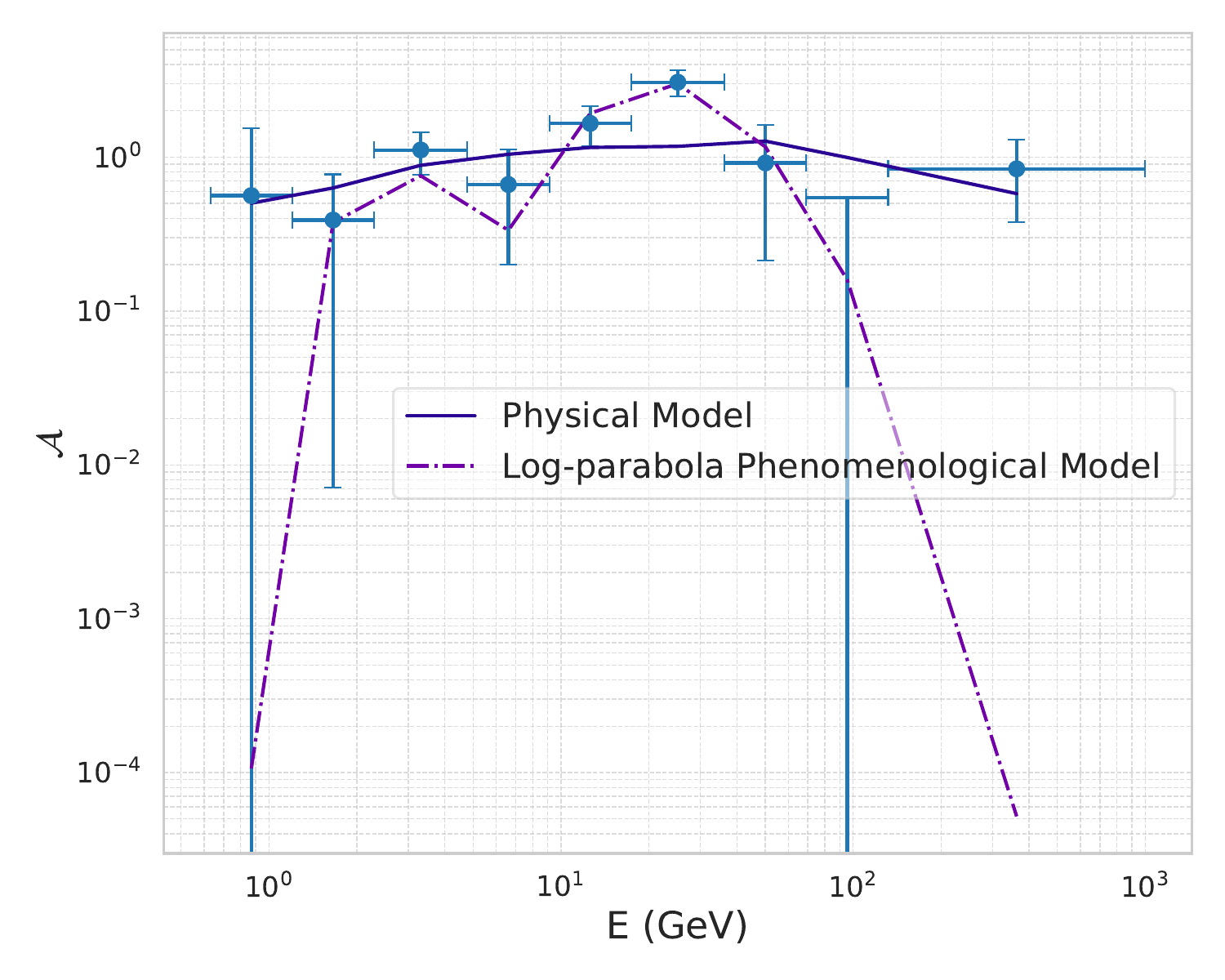}
    \includegraphics[width=0.45\textwidth]{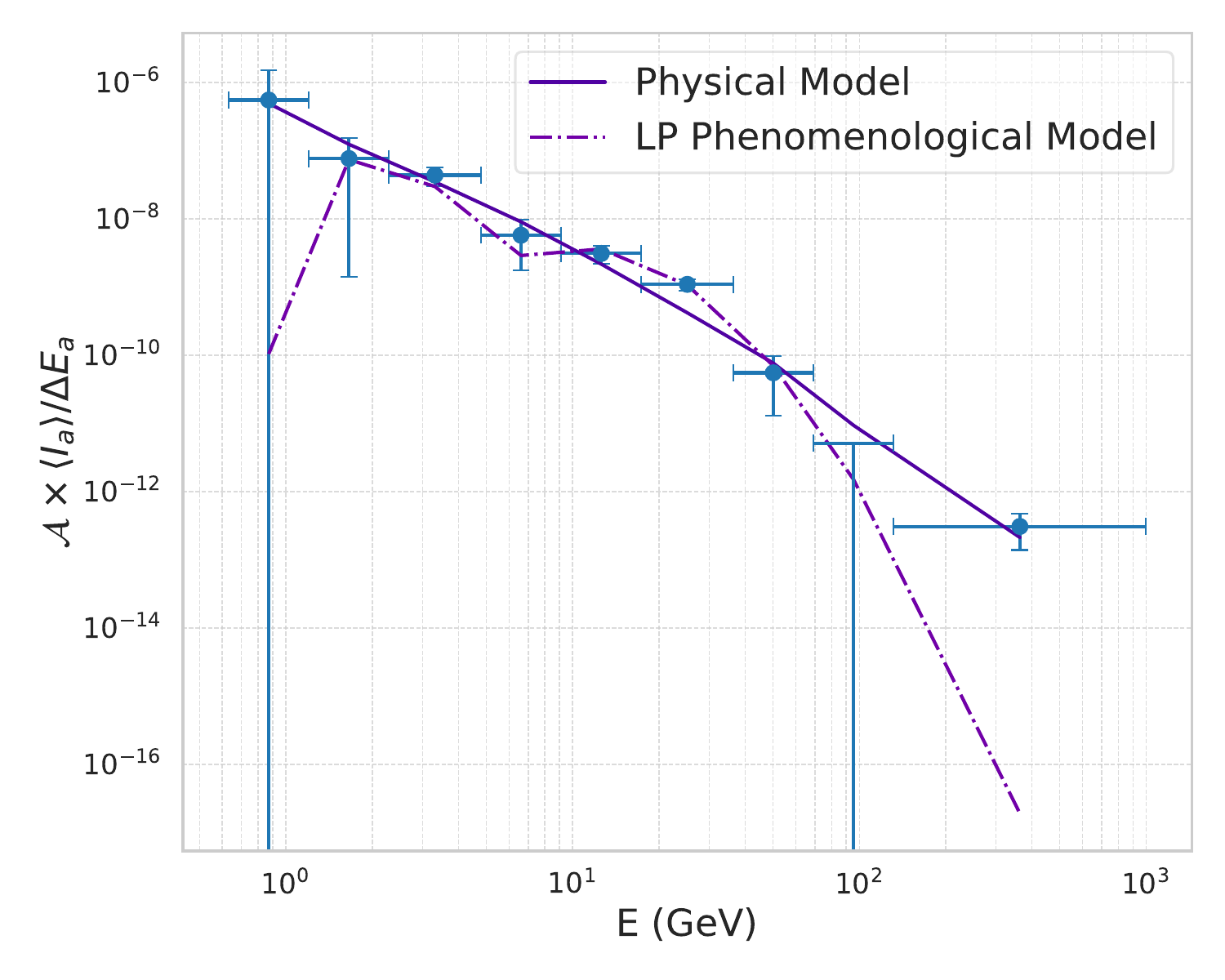}
    \caption{Top: Angular scale (left) and redshift (right) behaviour for LP (phenomenological) and Physical Models. Bottom left: Energy behaviour for the LP and Physical Models in terms of raw matched filter amplitude. While the physical model shows a near-flat behaviour compared to the naive (PL) model, the LP model shows a significant curvature at low-to-mid energies. Bottom right: Energy behaviour for the LP, and Physical Models in terms of the differential energy amplitude with respect to the PL . }
    \label{fig:E_z_thetaplotspheno}
\end{figure*}

\begin{figure*}[htbp!]
\centering
    \includegraphics[width=\textwidth]{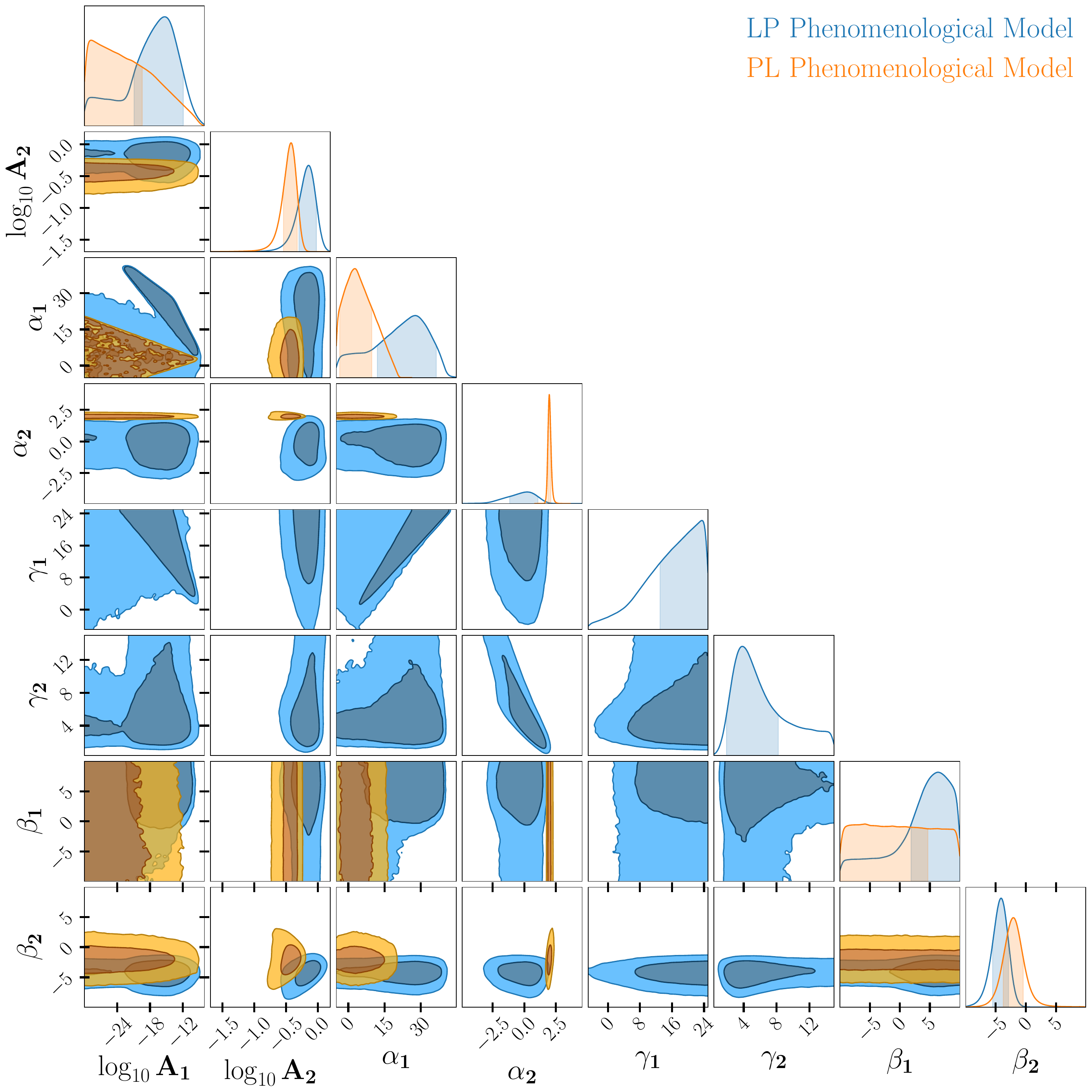}
    \caption{Left: Constraints on the parameters of the power-law (PL) and the log-parabola (LP) \pheno\ models. In all panels, the 2D contours refer to the 68\% and 95\% credible regions, with the shaded areas in the 1-D subplots denoting the 68\% credible interval for the associated posterior distributions. }
    \label{fig:contourplotspheno}
\end{figure*}

Parameter inference is conducted via Markov Chain Monte Carlo (MCMC) sampling using the affine-invariant ensemble sampler implemented in \texttt{emcee} \cite{foreman2013emcee}, and posterior distributions are visualized with \texttt{ChainConsumer} \cite{hinton2016chainconsumer}. The number of MCMC walkers is set to twice the number of free parameters in the model. We adopt uniform priors for all parameters and assume a Gaussian likelihood function.
For the spectral index $\mu_{\rm BLZ}$, we impose a prior range of $[1.5, 3.0]$, following the constraints reported in Ref.\ \cite{korsmeier2022flat}. All other physical and phenomenological parameters are assigned broad, non-informative priors. The convergence of the MCMC chains is evaluated by estimating the autocorrelation function and the corresponding autocorrelation time, as outlined in Ref.\ \cite{goodman2010ensemble}.

To qualitatively assess the measured signal, we examine its dependence on angular scale, photon energy, and redshift, as shown in Fig.\ \ref{fig:E_z_thetaplotspheno} for the measurements and the different models considered.  The angular dependence shown in the top-left panel is the mean of the cross-correlation 
across all energy and redshift bins.
The energy and redshift dependencies are derived using a matched-filter amplitude
\begin{equation}
{\cal A} = \frac{{\bm \Xi}^{\sf T}\,{\bm \Gamma}^{-1}\,\bar{\bm \Xi}_{\rm M}}{\bar{\bm \Xi}_{\rm M}^{\sf T}\,{\bm \Gamma}^{-1}\,\bar{\bm \Xi}_{\rm M}}\;,
\end{equation}
where ${\bm \Gamma}$ denotes the covariance matrix (see Appendix~\ref{sec:cov}) and $\bar{\bm \Xi}_{\rm M}$ corresponds to a simple model that approximately intercepts the expected behavior of the signal. In this case, we consider the best-fit PL model as the simple model. 
The vector ${\bm \Xi}$ represents either the estimator defined in Eq.\ (\ref{LSestimator_GalxGam}) in the case of measurements or the best-fit LP phenomenological and physical model counterparts. Throughout this section, boldface symbols denote either the full vector of the correlation function or the covariance matrix, as opposed to their individual components.\footnote{Look at \url{www.github.com/BhashinT/Matched_Filter_FermixDES.git} for a detailed derivation of the matched-filter signal-to-noise ratio.} 
%
The corresponding uncertainty on the matched-filter amplitude is given by
\begin{equation}
\sigma^2_{\cal A} = 
\frac{
({\bm \Gamma}^{-1}\,\bar{\bm \Xi}_{\rm M})^{\sf T}\,{\bm \Gamma}\,({\bm \Gamma}^{-1}\,\bar{\bm \Xi}_{\rm M})
}{
(\bar{\bm \Xi}_{\rm M}^{\sf T}\,{\bm \Gamma}^{-1}\,\bar{\bm \Xi}_{\rm M})^2
}\;,
\end{equation}
as shown in Fig.\ \ref{fig:E_z_thetaplotspheno} on the data estimates.


We find a clear positive detection, with the signal exhibiting a moderate to high curvature in its energy dependence, along with a mostly flat trend for the redshift dependence. The cross-correlation starts deviating from zero below approximately 100 arcmin. Since our signal is significant around this angular range, a PSF that is much smaller than the degree scale is crucial to properly sample the signal.

To quantitatively assess the statistical significance of the observed signal, we evaluate its deviation from the null hypothesis of pure noise using the three models introduced earlier and two complementary statistical approaches. 

First, we employ a $\Delta\chi^2$ test statistic, where the chi-squared is defined as  
\begin{equation}
\chi^2({\bm P}_{\rm mod}) =
\left[{\bm \Xi}_{\rm data} - {\bm \Xi}_{\rm th}({\bm P}_{\rm mod})\right]^{\sf T}
{\bm \Gamma}^{-1}
\left[{\bm \Xi}_{\rm data} - {\bm \Xi}_{\rm th}({\bm P}_{\rm mod})\right],
\end{equation}
with ${\bm \Xi}_{\rm data}$ denoting the data vector and ${\bm \Xi}_{\rm th}$ the theoretical prediction for a model described by the parameter set ${\bm P}_{\rm mod}$.  
We define the test statistic as $\Delta\chi^2_{\rm mod} = \chi^2_{\rm null} - \chi^2({\bm P}_{\rm mod}^\star)$, where $\chi^2({\bm P}_{\rm mod}^\star)$ corresponds to the minimum chi-squared obtained for the best-fit parameters ${\bm P}_{\rm mod}^\star$, and $\chi^2_{\rm null}$ represents the no-signal case, i.e.\ ${\bm \Xi}_{\rm th} = 0$.  

As an independent measure of detection significance, we also compute the matched-filter signal-to-noise ratio (SNR; see e.g.\ Ref.\ \cite{becker2016cosmic}), defined as  
\begin{equation}
{\rm SNR}({\bm P}_{\rm mod}) = 
\frac{
{\bm \Xi}_{\rm data}^{\sf T} {\bm \Gamma}^{-1} {\bm \Xi}_{\rm th}({\bm P}_{\rm mod})
}{
\sqrt{
{\bm \Xi}_{\rm th}^{\sf T}({\bm P}_{\rm mod}) {\bm \Gamma}^{-1} {\bm \Xi}_{\rm th}({\bm P}_{\rm mod})
}},
\end{equation}
and we report ${\rm SNR}_{\rm mod} \equiv {\rm SNR}({\bm P}_{\rm mod}^\star)$ evaluated at the best-fit model parameters.

Table~\ref{tab:chi2comp} summarizes the results of our detection significance analysis.  
Applying the \pheno\ framework to the complete data set reveals a clear cross-correlation signal, corresponding to $\mathrm{SNR}_{\rm mod}=7.85$ for the log-parabola case and $\mathrm{SNR}_{\rm mod}=5.70$ for the power-law model.  
For both model variants, their best-fit parameters derived from the maximum of the joint posterior (and their associated 68\% credible intervals obtained from the marginalized one-dimensional posteriors) are reported in Table~\ref{tab:bestfit_pheno}.  
The corresponding 68\% and 95\% credible contours are displayed in Fig.\ \ref{fig:contourplotspheno}. 

We note that the distribution of $A_1$ in the LP model shows a non-trivial behaviour which deserves some clarifications. The contours exhibit a mild multi-modality: $A_1$ is either very small (close to zero) with low $\alpha_1$ and $\gamma_1$ values, or larger with very high $\alpha_1$ and $\gamma_1$ values. In the first case (i.e, a very small $A_1$ with low $\alpha_1$ and $\gamma_1$ values), the 1-halo term is suppressed, while in the second case (large $A_1$ with very high $\alpha_1$ and $\gamma_1$ values), it is relevant at low energies. The two models yield comparable likelihood values, though the second is marginally preferable, with an $\mathrm{SNR}$ of $7.85$ compared to $7.76$ for the first case. The second case also fits better the energy behaviour, providing an additional peak at low-$E$ from the 1-halo term, see Fig.\ \ref{fig:E_z_thetaplotspheno}, while, on the other hand, a pure 2-halo correlation fits better the angular scaling.

To gain further insight into the origin and behaviour of the detected correlation, we repeat the statistical tests on data subsets defined by redshift, photon energy, and angular scale.  
Specifically, the Low-$z$ and High-$z$ samples correspond to the first three and last two \rmagic\ redshift bins, respectively, while Low-$E$ and High-$E$ subsets include energy bins below and above $9\,\mathrm{GeV}$ (the first four and last five bins, respectively); and Small-$\theta$/Large-$\theta$ regimes separate angular scales smaller or larger than the 68\% containment angle of the \Fermi\ PSF, as listed in Table~\ref{tab:enbins}.

The trends summarized in Table~\ref{tab:chi2comp} indicate that the statistical significance of the correlation signal is primarily driven by photons larger than 9 GeV, and mid-to-large angular separations, with the 2-halo phenomenological component dominant across the angular range. We also find  that the redshift behaviour is mostly flat across the spectrum in the power-law and physical models. The log-parabola model shows a slight preference for signals at low redshifts, which is consistent with large-scale structure formation. 
It is remarkable however that all subsets provide a non-negligible statistical significance.

The preference for mid-to-large angular scales suggests that the observed signal is not the product of a few luminous, extremely bright sources (such as individual BLZ which would contribute to smaller scales), but rather reflects the clustered distribution of numerous extragalactic emitters, or that of extended sources.  
At lower energies, the broad \Fermi\ point-spread function limits angular resolution and consequently dilutes the statistical significance.  
Conversely, the better angular resolution at higher energies allows a better characterization of the correlation signal.
The redshift and angular dependencies are broadly consistent with expectations for any \g-ray source class that follows the large-scale matter distribution.  

The spectral index of the two-halo component for the power-law model, $\alpha_2 = 1.91^{+0.07}_{-0.09}$, is noticeably harder than the mean unresolved \g-ray background spectrum ($\alpha \approx 2.3$, see Ref.\ \cite{2015ApJ...799...86A}).  
This spectral index is consistent with emission from BL Lac-type sources, which are expected to dominate the unresolved \g-ray population just below the \Fermi\ detection threshold. However, we can also see that the LP model, with an added curvature parameter, is significantly preferred over the PL model, the former being able to account for the excess in the energy spectrum at mid energies.

\begin{table}[]
\centering
\begin{tabular}{l|l|l|l|l}
\hline
\textbf{Parameter} & \textbf{$\mathbf{68\%}$ C.I. (PL)} & \textbf{Best fit (PL)} & \textbf{$\mathbf{68\%}$ C.I. (LP)} & \textbf{Best fit (LP)} 
        \\ \hline
        \hline
$\rm \log_{10} A_1$ & $[-29.93,-19.14]$& $-19.50$& $[-20.91,-11.92]$& $-17.44$\\ 
$\rm \log_{10} A_2$                 & $[-0.53,-0.32]$& $-0.39$& $[-0.29,-0.022]$& $-0.12$\\
$\alpha_1$             & $[-3.55,9.73]$& $3.13$& $[12.02,36.47]$& $28.82$\\ 
$\alpha_2$             & $[1.82,1.98]$& $1.91$& $[0.90,1.76]$& $0.57$\\ 
$\gamma_1$             & N.A.                             & N.A.                                 & $[13.01,24.92]$& $18.15$\\ 
$\gamma_2$             & N.A.                             & N.A.                                 & $[1.90,8.21]$& $3.24$\\ 
$\beta_1$              & $[-9.93,4.97]$& $9.59$& $[1.87,9.94]$& $6.99$\\ 
$\beta_2$              & $[-3.75,-0.44]$& $-2.48$& $[-5.50,-2.85]$& $-4.13$\\ \hline
\end{tabular}
\caption{The $68\%$ credible interval (C.I.) and global best fit values of the parameters for the power-law and log-parabola \pheno\ models, denoted as PL and LP respectively. }
\label{tab:bestfit_pheno}
\end{table}

\subsection{Physical Interpretation \label{subsec:physical_interpretation_galxgam}}

Having established the presence of a statistically significant cross-correlation signal, we now aim to study its underlying astrophysical $\gamma$-ray components. Different source populations are expected to produce distinct cross-correlation signatures as functions of angular separation, photon energy, and redshift. As discussed above, we consider blazars to be the dominant population contributing to the UGRB and to have intrinsic angular sizes much smaller than the \textit{Fermi}-LAT PSF.
Consequently, the angular correlation function of the 1h term is effectively determined by the instrument PSF. However, as indicated by Table~\ref{tab:chi2comp}, the data requires a cross-correlation extending to larger angular scales, and the statistical significance of the detection is in fact driven by the 2h component. This implies that our measurement is sensitive to the large-scale clustering of blazars.

To investigate the blazar properties capable of reproducing the observed cross-correlation signal, we perform the statistical tests described in the previous section, now using a physical model that includes both 1h and 2h contributions to the cross-correlation between galaxies and $\gamma$-ray emission from blazars, as formulated in Eq.\ (\ref{eq:physmdl}).

Ref.\ \cite{korsmeier2022flat} found that two blazar subpopulations are needed to describe the angular power spectrum (APS) measure in Ref. \cite{Ackermann:2018wlo}, with Flat Spectrum Radio Quasars (FSRQs) dominating at GeV energies and BL Lacertae (BL Lacs) becoming dominant above a few GeV. As discussed earlier, our first two energy bins yield inconclusive results with limited constraining power. Therefore, in this work we focus our modeling on a unified BLZ description, tailored on the BL Lac population.

\begin{figure*}[htbp!]
\centering
    \includegraphics[width=0.75\textwidth]{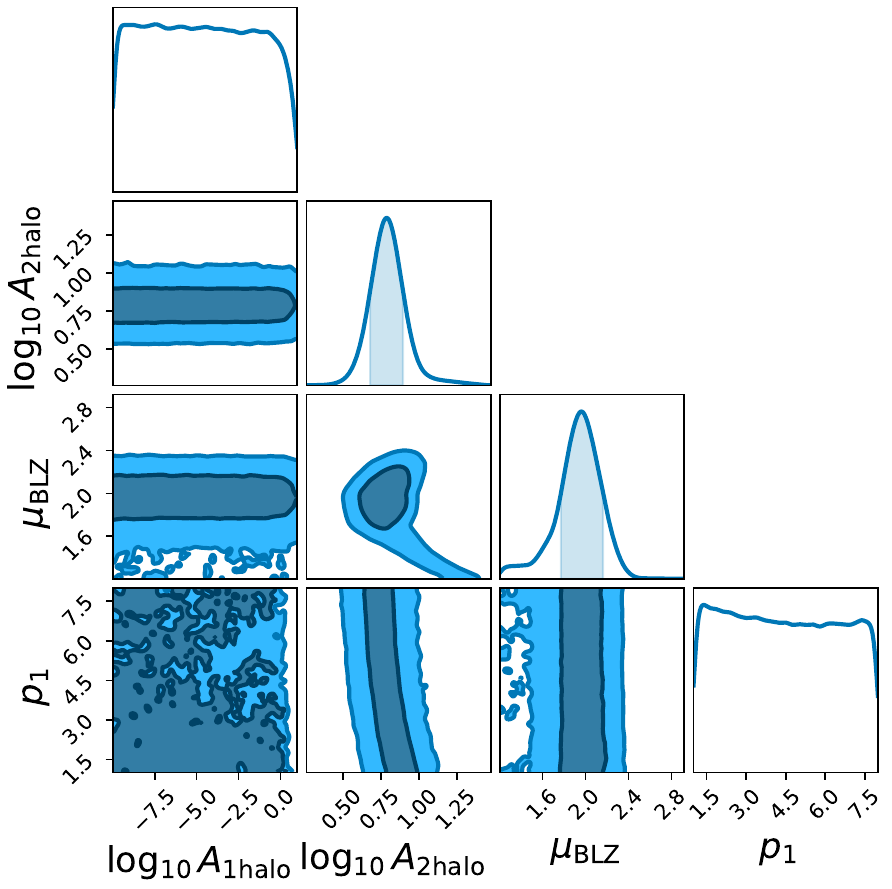}
    \caption{Constraints on the parameters describing the physical model BLZ, based on the
BL Lac model from Ref.\ \cite{korsmeier2022flat}, and assuming a single population accounting for the measured cross-
correlation. As before, the 2D contours refer to the 68\% and 95\% credible regions, with
the shaded areas in the 1D subplots denoting the 68\% credible interval for the 1D posterior
distribution.}
    \label{fig:contourplotsphys}
\end{figure*}

We show the constraints on the physical model in Fig.\ \ref{fig:contourplotsphys}. As already discussed in Eq.\ (\ref{eq:physmdl}), the normalizations of the 1h and 2h components are treated as independent free parameters, to account both for potential deviations in the overall GLF amplitude and for possible variations in the relation between host-halo mass and $\gamma$-ray luminosity, which would differently affect the linear and non-linear bias.
As found also for the phenomenological models, the correlation is dominated by the 2h component, see Fig.\ \ref{fig:E_z_thetaplotspheno}. The 1h term was expected to be low because \rmagic\ galaxies have low star formation rates and thus it is unlikely they host blazars.

Overall, we find that the BLZ model returns a SNR of 5.89, which is similar to the SNR obtained from the PL phenomenological model. This is explained by both the redshift and energy behaviour of the BLZ-only model being similar to that of the phenomenological model.
However, as also found in Ref. \cite{thakore2025high}, we noticed  that the curvature of the LP model seems to be highly preferred with $\Delta \chi^2=\chi^2_{PL}-\chi^2_{LP} \sim 22$.
This might indicate that while BLZ sources are an important component to the UGRB, additional physical contributions to the energy spectrum might be required to explain the cross-correlations, either astrophysical (such as misaligned Active Galactic Nuclei (mAGNs) or SFGs) or exotic sources (such as DM). Or alternatively, a different BLZ model has to be introduced. 
A complete inspection of the origin of this deviation in the energy spectrum shall be explored in an upcoming work.

\begin{table}[]
\centering
\begin{tabular}{l|l|l}
\hline
\textbf{Parameter} & \textbf{$\mathbf{68\%}$ C.I.} & \textbf{Best-fit} \\ \hline
\hline
$\log_{10}A_{1\rm halo}$                & $[-\infty,-2.78]$   & $-3.09$       \\ 
$\log_{10}A_{2\rm halo}$                  & $[0.68,0.90]$      & $0.88$          \\ 
$\mu_{\rm BLZ}$                & $[1.77,2.15]$       & $1.96$          \\ 
$p_1$                 &  Unconstrained         & $1.24$                    \\ \hline
\end{tabular}
\caption{The $68\%$ credible interval and global best fit values of the \phys\ model parameters of Eq. (\ref{eq:physmdl}). }
\label{tab:bestfit_phys}
\end{table}

\begin{figure*}[htbp!]
\centering
    \includegraphics[width=0.65\textwidth]{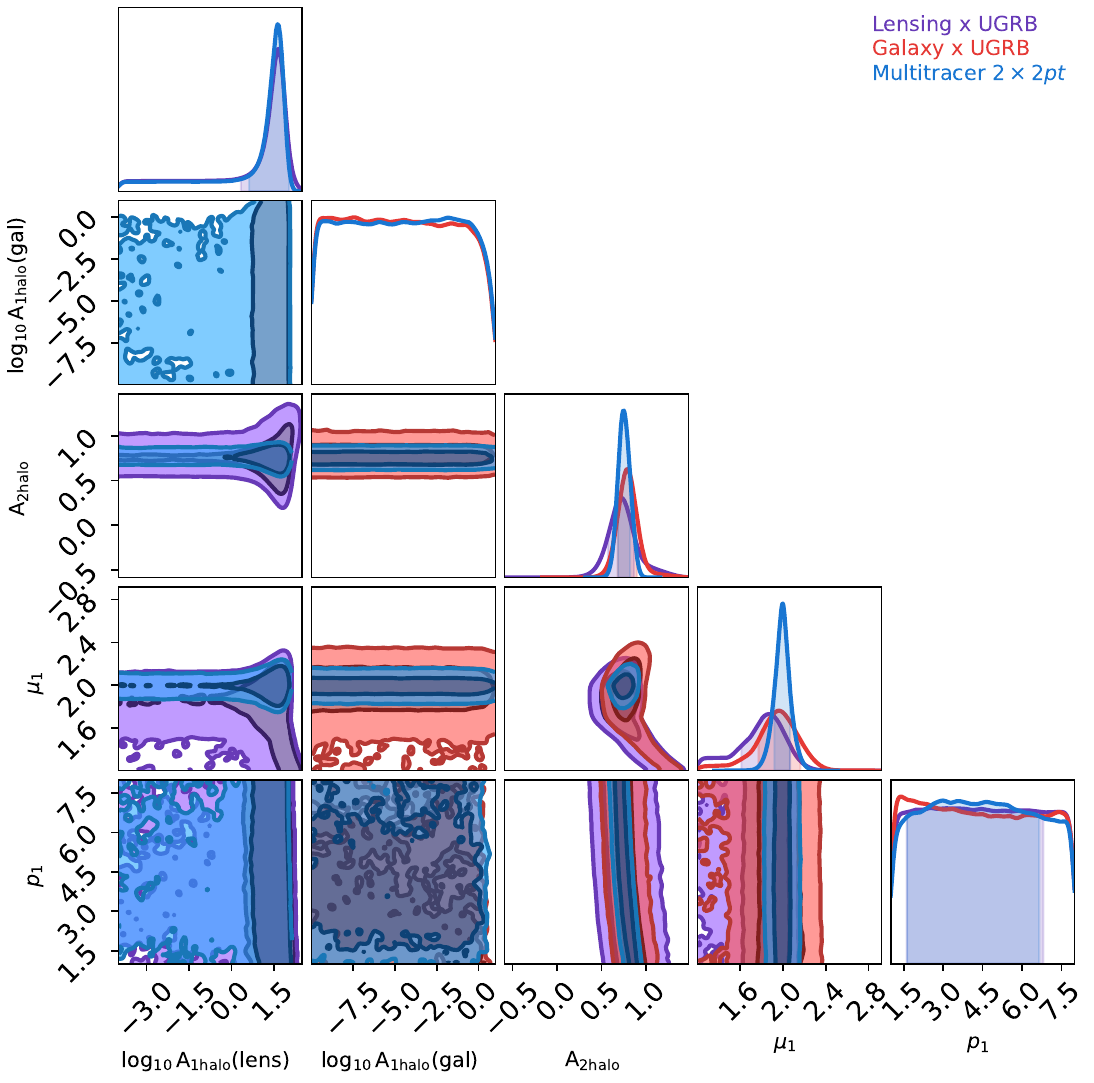}
    \caption{Joint parameter-space contours for the astrophysical model, combining constraints from multiple tracers.}
    \label{fig:contour_plot_multitracer}
\end{figure*}

\section{Results from a Multi-Tracer Approach \label{sec:multitracer_results}}
In this section, we combine the analysis discussed in the previous sections with the study of cross-correlations between \g-rays and weak lensing of Ref.\ \cite{thakore2025high}. Namely, we extend our analysis to a coherent $2\times2$pt framework that jointly models galaxy–\g-ray and lensing–\g-ray cross-correlations. This multi-tracer approach explicitly accounts for the shared underlying large-scale structure. Because each gravitational probe weights the UGRB through a distinct kernel, their combination provides a more informative, effective characterization of the \g-ray field associated with large-scale structure. Considering the cross-covariances between observables yields tighter and more robust parameter constraints, than analysing each cross-correlation separately.
\\
In order to carry out such a multi-tracer approach we concatenate the datasets and covariances for both cross-correlations, and generate cross-covariances for the galaxy and lensing components to capture the off-diagonal elements of the combined covariance (for more details, see Appendix \ref{sec:cov}). The combined covariance can then be written as:
\[
\mathbf{C}_{e_t \times \gamma;\delta_g\times\gamma}
=
\begin{bmatrix}
C_{e_t\times \gamma} & C_{e_t \times \delta_g} \\
C_{e_t \times \delta_g} & C_{\delta_g\times \gamma}
\end{bmatrix},
\]
where $e_t$, $\delta_g$, and $\gamma$ denote the tangential ellipticity, galaxy overdensity, and the the \g-ray intensity components that contribute to the covariance. 

Concerning the interpretation, we consider only the physical model, which allows to calculate all the correlations signals in a unifed way, contrary to the phenomenological model, for which we have to define a single model for each cross-correlation. The physical model is the one described in Section~\ref{subsec:physical_interpretation_galxgam} and Ref.\ \cite{thakore2025high}, and it is composed only by BLZ for the \g-ray part.

Figure \ref{fig:contour_plot_multitracer} compares the parameter constraints obtained from the single-tracer analyses with those from the combined multi-tracer approach. We assign different $A_{\rm 1halo}$ terms for the lensing and galaxy cross-correlations, owing to the fact that this factor corrects the modelling of the blazar occupation of halos in the first case and of the blazar occupation of \rmagic\ galaxies in the second, something that could be very different, given the specific properties of the \rmagic\ sample.

Note that
the results obtained from lensing are well compatible with those obtained from galaxies, and show some level of complementarity. Overall, the multi-tracer framework tends to yield constraints that are stronger. In particular, 
the multi-tracer analysis provides noticeably tight bounds on the normalisation amplitude and $A_{\rm 2halo}$ and on the spectral index $\mu_1$. 
The redshift-evolution parameter $p_1$ remains only weakly constrained or unconstrained in all configurations. The overall sensitivity to redshift evolution remains limited, suggesting that current data are not yet sufficient to robustly determine it.

\begin{table}
\centering
\begin{tabular}{lc|ll|ll|lll}
\hline
 & \multicolumn{7}{c}{Subsamples Selected (Galaxy $\times$ \g-rays)} \\
 \cline{2-8}
 & Full & Low-$z$ & High-$z$ & Low-$E$ & High-$E$ & Small-$\theta$ & Large-$\theta$ & Combined\\
\hline
\hline
$\Delta \chi^2_{\rm BLZ}$ & $34.15$ & $15.38$ & $20.17$ &  $8.41$ & $25.93$ & $12.79$ & $34.39$ & $112.07$\\
${\rm SNR}_{\rm BLZ}$ & $5.86$ & $3.92$ & $4.62$ & $2.99$ & $5.29$ & $3.58$ & $5.86$ & $10.31$\\
 \hline
\end{tabular}
\begin{tabular}{lc|ll|ll|lll}
\hline
 & \multicolumn{7}{c}{Subsamples Selected (Lensing $\times$ \g-rays)} \\
 \cline{2-8}
 & Full & Low-$z$ & High-$z$ & Low-$E$ & High-$E$ & Small-$\theta$ & Large-$\theta$ & Combined\\
\hline
\hline
$\Delta \chi^2_{\rm BLZ}$ & $41.01$ & $1.71$ & $41.83$ & $1.31$ & $39.74$ & $0.05$ & $45.45$ & $112.07$\\
${\rm SNR}_{\rm BLZ}$ & $6.45$ & $1.43$ & $6.50$ & $2.33$ & $6.32$ & $0.96$ & $6.80$ & $10.31$\\
 \hline
\end{tabular}
\caption{$\Delta \chi^2_{\rm mod}$ and ${\rm SNR}_{\rm mod}$ computed for the \phys\ models comprising of Blazars, using either the full data set or the various subsamples discussed in the text. For the Low-$z$ case we selected the first two redshift bins , while for the High-$z$ case the last three bins; the Low-$E$ subset refers to the first four energy bins, i.e., energies below 9 GeV, while the High-$E$ to the bins at higher energies; finally, the {Small-$\theta$/Large-$\theta$} cases correspond to data points below/above the $68\%$ containment angle of the \Fermi\ PSF. }  
\label{tab:chi2comp_multitracer}
\end{table}

Table \ref{tab:chi2comp_multitracer} summarises the statistical evidence of the multi-tracer results. The SNR is 10.3 for the BLZ physical scenario considered here. 
Such a high significance, combined with the compatibility of the results for the two tracers, allows us to undoubtedly conclude that the bulk of the UGRB is of extragalactic origin and correlates with LSS.

Table \ref{tab:chi2comp_multitracer} reports the significances of the combined best-fit model for subsets of the individual datasets.
Overall, the multi-tracer findings are consistent with the key results of Ref.\ \cite{thakore2025high}, with the significance of the weak-lensing--$\gamma$-ray signal being dominated by contributions from high redshifts, high energies, and large angular scales. We find that the statistical evidences for the individual galaxy--\g-ray and weak lensing--\g-ray cases (with the latter shown in Ref.\ \cite{thakore2025high}) mirror the behaviour shown in the multi-tracer analysis, with the galaxy--\g-ray cross-correlations depicting a preference for high energies and large angular scales while displaying a flat redshift behaviour, while the weak lensing case shows a preference for high redshifts, high energies, and large angular scales.
In both cases, the individual SNR values obtained from the multi-tracer best-fit are slightly lower than, although still comparable to, those from the the best-fits of the corresponding single-tracer analyses. This small reduction is clearly expected because the best-fit parameters, reported in Table \ref{tab:bestfit_phys_multitracer}, refer to a global joint analysis across tracers rather than being tailored to a single cross-correlation channel. 
The key point is that the combined SNR is higher than the individual ones and that the two individual findings are compatible with each other.

\begin{figure*}[htbp!]
\centering
    \includegraphics[width=0.45\textwidth]{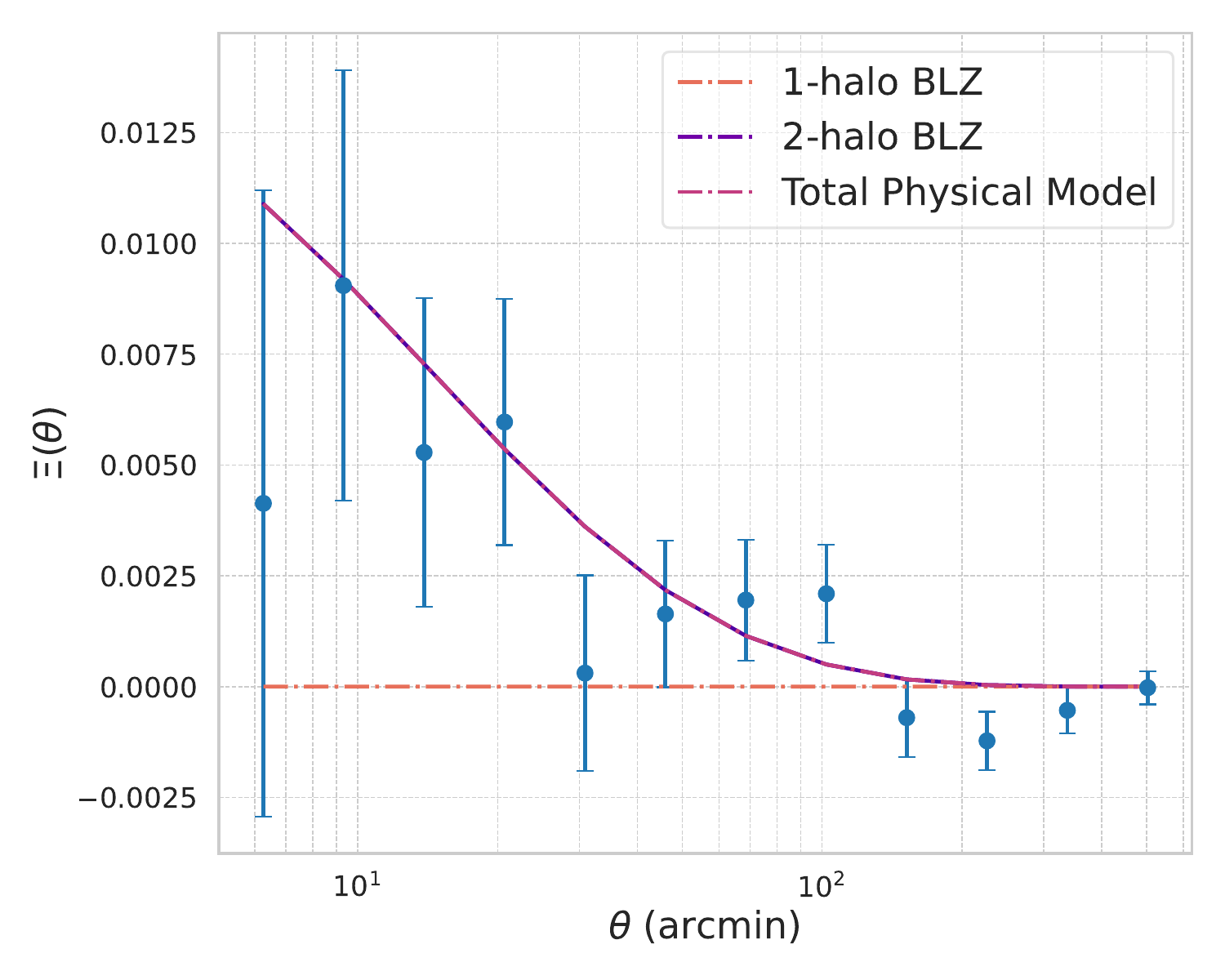}
    \includegraphics[width=0.45\textwidth]{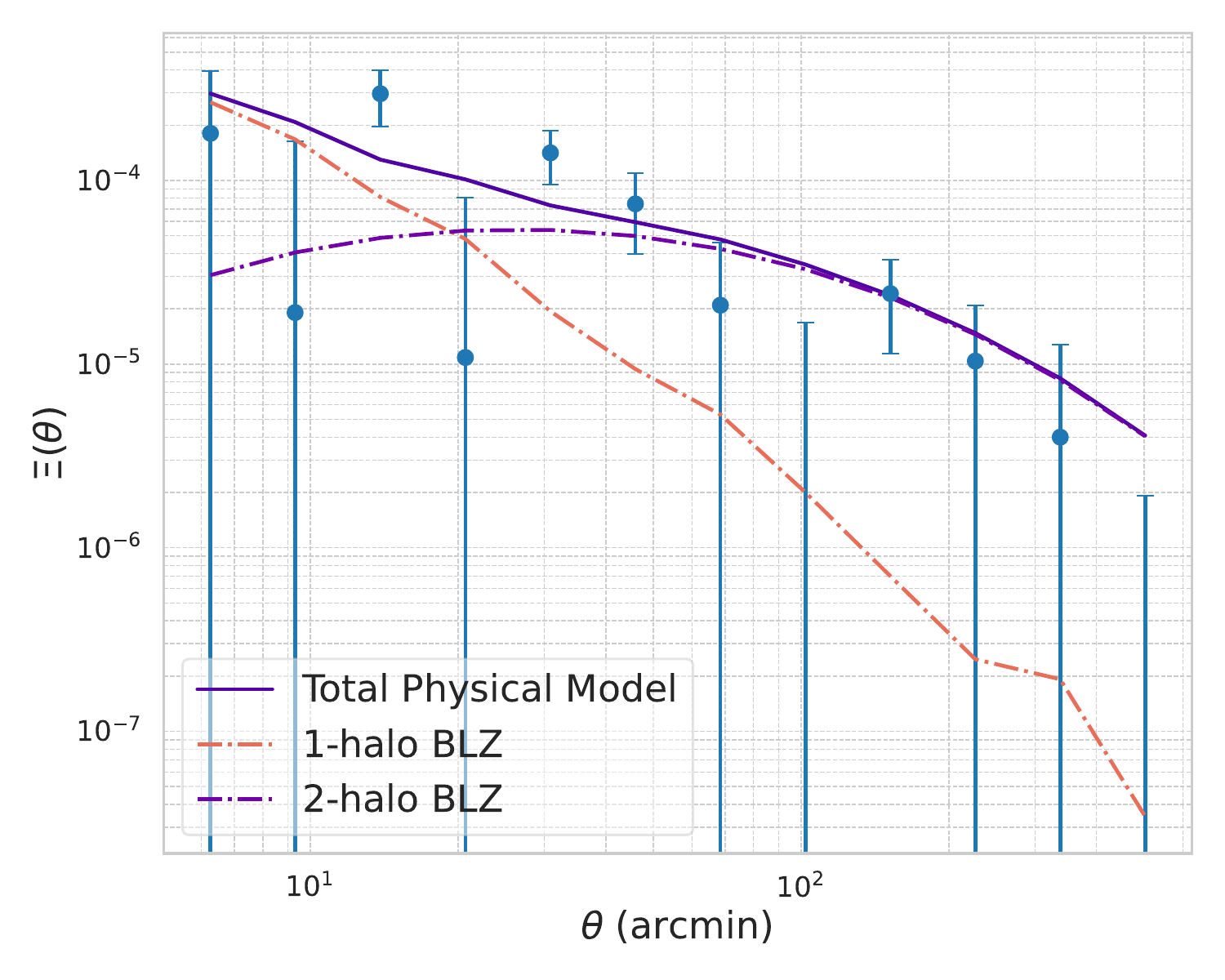}
    \caption{Multi-tracer results for galaxy-\g-ray (left) and lensing-\g-ray (right) cross-correlations, depicting their behaviour across the angular scales. }
    \label{fig:E_z_thetaplots_multitracer}
\end{figure*}

From the angular behaviour of the cross-correlations for the combined best-fit reported in Fig.\ \ref{fig:E_z_thetaplots_multitracer}, we can appreciate a 2-halo dominance emerging in both analyses. This tells us that we actually measured a LSS correlation between \g-ray sources and matter tracers.

\begin{table}[]
\centering
\begin{tabular}{l|l|l}
\hline
\textbf{Parameter} & \textbf{$\mathbf{68\%}$ C.I.} & \textbf{Best-fit} \\ \hline
\hline
$\log_{10} A_{1\rm halo}$ (lensing)                & $[0.62.,2.04]$& $1.66$\\ 
$\log_{10} A_{1\rm halo}$ (galaxies)                & $[-\infty,-2.67]$& $-2.36$\\
$A_{2\rm halo}$                  & $[0.68,0.82]$& $0.72$\\ 
$\mu_{\rm BLZ}$                & $[1.91,2.06]$& $1.99$\\ 
$p_1$                 & $[1.62,6.65]$& $4.42$\\ \hline
\end{tabular}
\caption{The $68\%$ credible interval and global best fit values of the \phys\ model parameters of Eq. (\ref{eq:physmdl}). }
\label{tab:bestfit_phys_multitracer}
\end{table}

\section{Conclusions} \label{sec:concl}

In this paper, we have investigated the connection between large-scale structure and the unresolved $\gamma$-ray background using both single-tracer and multi-tracer cross-correlation analyses. 

Using galaxy clustering as a tracer of the matter distribution, we measured a cross-correlation between DES \rmagic\ galaxies and the \Fermi\ $\gamma$-ray sky at high statistical significance.
The observed signal is dominated by the two-halo large scale angular correlation, and shows a curved energy spectrum and a relatively flat redshift behaviour. A phenomenological model with a log-parabolic energy spectrum is preferred with respect to a power-law one. From a physical standpoint, we fit the measurement considering blazars as the dominant \g-ray source population. In our model, the BLZ spectrum is a power-law and thus it does not reproduce the data to the same degree as the phenomenological log-parabolic spectral model. This might indicate that an additional astrophysical contribution or a different BLZ model description may be required. A dedicated exploration of these possibilities, including alternative source populations (possibly non-standard, such as DM) will be undertaken in a follow-up work. 

In the second part of the paper, we implemented a multi-tracer framework that combines galaxy–\g-ray and weak-lensing-\g-ray cross-correlations, and explored consequences for a blazar-only astrophysical model. 
This joint analysis yields tighter bounds on the normalisation parameters and energy spectral index, demonstrating the advantages of coherently modelling the shared large-scale structure and cross-covariances across tracers. 
The two single-tracer approaches show compatible results, and their combination in the multi-tracer analysis leads to an evidence for a physical correlation with an SNR of to 10.3.

This means that it is now understood that the UGRB has an extragalactic origin and correlates with the large-scale structure of the Universe.
The cross-correlation technique is now mature to be used for characterising the composition and evolution of the UGRB.


\acknowledgments
BT, MR and NF acknowledge support from the  Research grant TAsP (Theoretical Astroparticle Physics) funded by \textsc{infn}. BT acknowledges the support provided by the PNRR grant ex DM 118 scholarship. The work of MR and NF is supported by the Italian Ministry of University and Research (\textsc{mur}) via the PRIN 2022 Project No. 20228WHTYC – CUP: D53C24003550006. SC acknowledges support from the Italian Ministry of University and Research (\textsc{mur}), PRIN 2022 `EXSKALIBUR – Euclid-Cross-SKA: Likelihood Inference Building for Universe's Research', Grant No.\ 20222BBYB9, CUP D53D2300252 0006, Grant No.\ ZA23GR03. BT, MR, SC, and NF acknowledge support from the European Union -- Next Generation EU.

\bibliographystyle{unsrt}
\bibliography{bibliography}
\appendix
\section{Covariance Matrix \label{sec:cov}}
\subsection{Covariance Matrix for Galaxy-\g-ray Cross-Correlations \label{subsec:galgam_cov}}
The covariance matrix used for the galaxy-\g-ray cross-correlations follows an analytical approach. In the Gaussian approximation, we can express the covariance in harmonic space as:
\be
\widehat\Gamma_{ar\ell,bs\ell^\prime} = \frac{ \delta^{\rm K}_{\ell\ell^\prime}}{(2\ell+1)\Delta \ell f_{\rm sky} }\left[C_\ell^{ar}C_{\ell^\prime}^{bs}+ \big(C_{\ell^\prime}^{rs}+ \nn^{rs}\big) \big(C_\ell^{ab} + \nn^{ab}\big)\right],
\label{eq:covCl_galxgam}
\ee
where, as in the main text, the redshift bins for the shear are indicated with indices $r,s$ while energy bins for the UGRB are denoted by indices $a,b$. The $C_\ell$'s denote (auto and cross) angular power spectra and $\cal N$ the noises. For \rmagic\ lens galaxies, the latter is taken from the shot noise given in Ref. \cite{prat2022dark}, i.e., [0.022,0.038,0.058,0.029, and 0.025] $\rm gal/arcmin^2$ for each of the five redshift bins respectively. In order to include a more realistic prediction for the angular power spectrum of galaxies that accounts for the masking effects from the DES sky coverage, we convolve a mode-coupling matrix (MCM) with the theoretical DES Y3 auto/cross-power spectra such that $\tilde{C_\ell^{rs}} = \mathcal{M} C_{\ell}^{rs}$, where $\mathcal{M}$ is the MCM estimated using \texttt{Namaster} (see e.g. Refs.\cite{alonso2019unified,garcia2019disconnected}\footnote{\url{https://github.com/LSSTDESC/NaMaster}}). It incorporates the coupling between the multipoles of the true power spectrum $C_{\ell}^{rs}$ caused by the mask, giving us the biased pseudo-power spectrum $\tilde{C_\ell^{rs}}$. We normalize this bias using the $f_{\rm sky}$ value estimated from the \rmagic\ lens galaxy mask \footnote{Provided at \url{https://des.ncsa.illinois.edu/releases/y3a2/Y3redmagic}.}. This procedure was shown in Ref. \cite{nicola2021cosmic} to yield a better accuracy for the calculation of covariance matrices.
\footnote{See also the \texttt{Namaster} tutorials at \url{https://namaster.readthedocs.io/en/latest/3Covariances.html}}

On the other hand, at the angular scales of interest, the \Fermi\ signal is dominated by Poisson noise~\cite{Ackermann:2018wlo}, thus in the case of the \g-ray angular power spectra, the MCM correction is negligible.

\begin{figure}
    \centering
    \includegraphics[width = \linewidth]{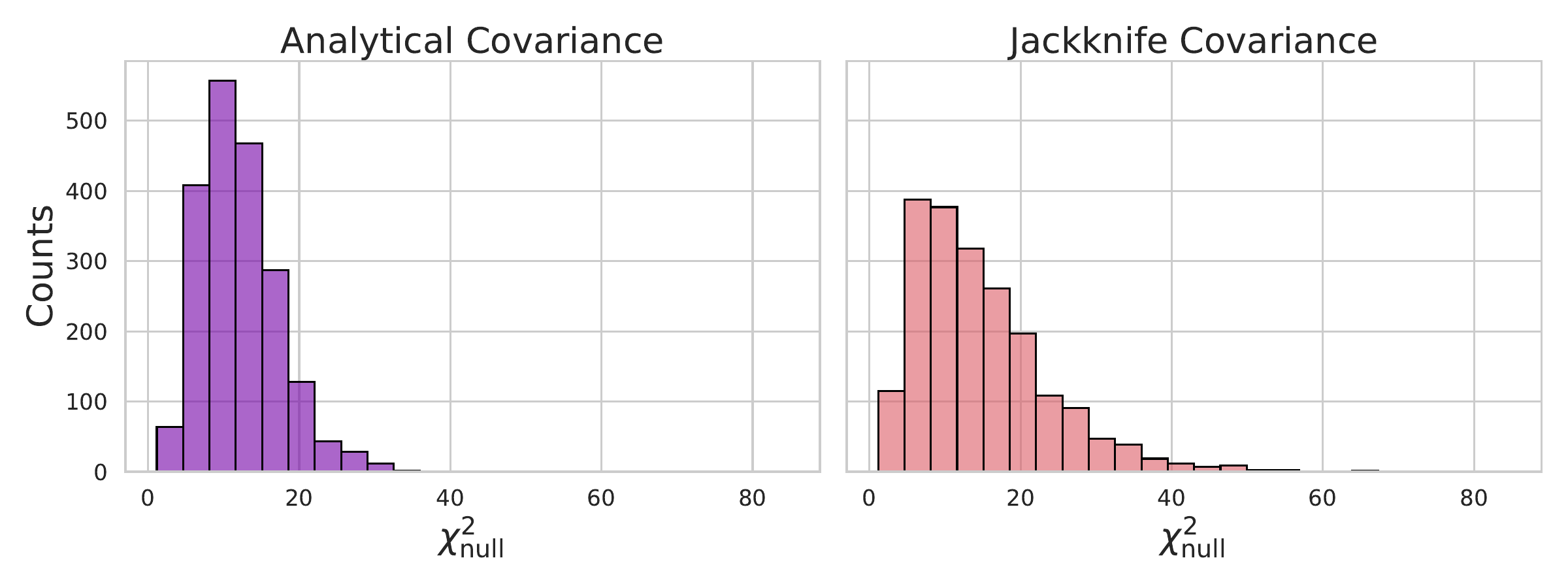}
    \caption{Chi-squared values between the (jackknife-rescaled) analytical covariance (left) and the jackknife covariance (right). As expected, the analytical covariance shows a more conservative distribution in response to the null datasets generated by the jackknife covariance.}
    \label{fig:chi_squared_comparison}
\end{figure}

All theoretical components involving $\gamma$-rays are corrected for the \Fermi\ PSF beam function before being included in the covariance calculation. The noise terms, $\nn$, are assumed to have no angular dependence. The factor $f_{\rm sky}$ accounts for the fraction of sky observed. For the cross-correlation analysis, we adopt a conservative approach by using only the overlapping region of the DES and \Fermi\  footprints. This is done by superimposing the DES and \Fermi\ \texttt{HEALPix} masks, and using the resultant sky fraction. When the two energy bins (indexed by $a$ and $b$) differ, we take the smaller of the two  $f_{\rm sky}$ values. The resultant harmonic covariance is then converted to the physical space by using the Legendre transformation described in Eq. (\ref{eq:Cl2xi}). 
The correction by $f_{\rm sky}$ is an approximate way to account for partial sky coverage.
In order to improve our estimate and also to reduce sensitivity to the specific setup used in generating the analytical covariance, we combine the covariance obtained analytically, with the uncertainties derived from jackknife resampling \footnote{While jackknife (and other internally estimated) covariance matrices might be biased for precise parameter inference, they are still useful for calibration. Normalising an analytical or simulated covariance with jackknife estimates offers a practical way to obtain a more realistic approximation to the true covariance.}  (see eg. Refs. \cite{sanchez2016cosmic}, \cite{crocce2016galaxy}, and \cite{prat2022dark} for more examples on combining two different types of covariances). We rescale the analytical covariance using the diagonal elements of the internally estimated jackknife covariance as:
\[
\rm{Cov}^{\rm comb}_{\theta_i,\theta_j}
= \rm{Corr}^{\rm Analytical}_{\theta_i,\theta_j}\,
\sigma^{\rm{JK}}_{\theta_i}\,
\sigma^{\rm{JK}}_{\theta_j},
\]
where $\mathrm{Corr}$ denotes the correlation matrix.
In order to test the veracity of this covariance, we generate 2000 null datasets at the highest energy and redshift bin combination using the associated jackknife covariance generated using \texttt{TreeCorr}, and test the $\chi^2$ values for the null hypothesis generated by both the pure jackknife and the (jackknife-rescaled) analytical covariances. The results are shown in Fig.\ \ref{fig:chi_squared_comparison}.  
We find that the distribution of the null $\chi^{2}$ obtained with the analytical covariance is squeezed around lower values than the one obtained with the jackknife covariance which has a larger tail. This indicates that the analytical covariance is more conservative, i.e.\ it predicts larger uncertainties for the same zero residual vector. The higher $\chi^{2}$ values from the jackknife estimate are expected: jackknife covariance matrices can be biased for finite numbers of jackknife regions, underestimating the true covariance. 

We thus use the (jackknife-rescaled) analytical covariance matrix for both the stand-alone as well as the multi-tracer analyses.

\subsection{Covariance Matrix for the Multi-tracer Analysis\label{subsec:mulitracer_cov}}
As mentioned in Section \ref{sec:multitracer_results}, the covariance is structured as:
\[
\mathbf{C}_{e_t \times \gamma;\delta_g\times\gamma}
=
\begin{bmatrix}
C_{e_t\times \gamma} & C_{e_t,\gamma \times \delta_g,\gamma} \\
C_{e_t,\gamma \times \delta_g,\gamma} & C_{\delta_g\times \gamma}
\end{bmatrix},
\]
where $e_t$, $\delta_g$, and $\gamma$ denote the tangential ellipticity, galaxy overdensity, and the the \g-ray intensity components that contribute to the covariance. In order to make the entire combined covariance analytical in nature (and thereby eliminate the need to use the Hartlap factor for just one component of the covariance), we replace the simulated covariance from Ref. \cite{thakore2025high} with the analytical version.
Similar to the analytical covariance in the previous subsection, we also normalise this covariance with its internally estimated jackknife covariance. 
We find here that the null signal chi-squared test for the cross-shear is $ \chi^2_{\rm null} \sim 467$, which aligns well with our expectations 
for the cross-component, which should be compatible with a vanishing signal. We also find that the log-parabola phenomenological model outputs an SNR for the weak lensing-\g-ray cross-correlations that drops from 8.9 found in Ref.\ \cite{thakore2025high} with the simulated covariance to $\sim 7.92$ in the mixed covariance. This can be attributed to the underlying conservative $f_{\rm sky}$ assumption taken for the shape noise, which, combined with the normalisation using the diagonals of the jackknife covariance, inflates the covariance beyond that of the simulations that were used to generate the shape noise in the previous work. Additionally, the use of the DR3 data for \Fermi instead of DR2 in Ref.\ \cite{thakore2025high} also contributes to a reduction in the SNR.
Despite the reduction in the SNR, we find that the physical inferences are consistent with that of Ref. \cite{thakore2025high}. As also shown in Table \ref{tab:chi2comp_multitracer}, we keep finding a higher significance for high energies, redshifts, and angular scales.  

The off-diagonal covariance which encodes the covariance between the lensing-\g-ray and clustering-\g-ray correlations can be written in harmonic space as:
\be
\widehat\Gamma_{ar\ell,bs\ell^\prime} = \frac{ \delta^{\rm K}_{\ell\ell^\prime}}{(2\ell+1)\Delta \ell f_{\rm sky} }\left[C_\ell^{ar}C_{\ell^\prime}^{bs}+ \big(C_{\ell^\prime}^{rs}\big) \big(C_\ell^{ab}\big)\right],
\label{eq:covCl_offidiagonal}
\ee
where $r$ and $s$ now represent the redshift bins of the tangential ellipticity and the galaxy overdensity, respectively, while $a$ and $b$ denote again the energy bins for the UGRB. \footnote{In order to calculate the weak lensing power spectrum, we use \textsc{PyCCL}, described in Ref.
\cite{chisari2019core} (with the code at \url{https://github.com/LSSTDESC/CCL}), following the prescription provided in Ref.\ \cite{faga2025dark}.} Considering that the lensing component is a spin-2 field, we then map the harmonic covariance to the physical space as described in Ref.\ \cite{friedrich2021dark}:
\be
\mathrm{Cov}\!\left[\hat{\Xi}^{ar}(\theta_1),\, \hat{\Xi}^{bs}(\theta_2)\right]
= \sum_{\ell,\ell'}
\frac{(2\ell_1+1)(2\ell_2+1)}{(4\pi)^2}\,
P_{\ell}^{2,ar}(\cos\theta_1)\,
P_{\ell'}^{bs}(\cos\theta_2)\,
\mathrm{Cov}\!\left[\hat{C}_{\ell}^{ar},\, \hat{C}_{\ell'}^{bs}\right],
\ee
where $P_{\ell}^{2,ar}(\cos\theta_1)$ denotes the associated Legendre polynomial of order two while $P_{\ell'}^{bs}(\cos\theta_2)$ represents the Legendre polynomial, to account for the spin-2 and spin-0 nature of the weak lensing and galaxy clustering fields, respectively. Since the physical space contains 432 measurements in total (12 angular bins $\times$ 9 energy bins $\times$ 4 redshift bins) for the lensing-\g-ray cross-correlations and 540 bins (12 angular bins $\times$ 9 energy bins $\times$ 5 redshift bins) for the galaxy clustering-\g-ray cross-correlations, the off-diagonal matrix resulted in a shape of $432 \times 540$ entries, while the complete multi-tracer covariance has a shape of $972 \times 972$ entries. 

\section{Assessing the Impact of the Luminosity Scaling of the Blazar GLF}
\begin{figure}
    \centering
    \includegraphics[width=0.5\linewidth]{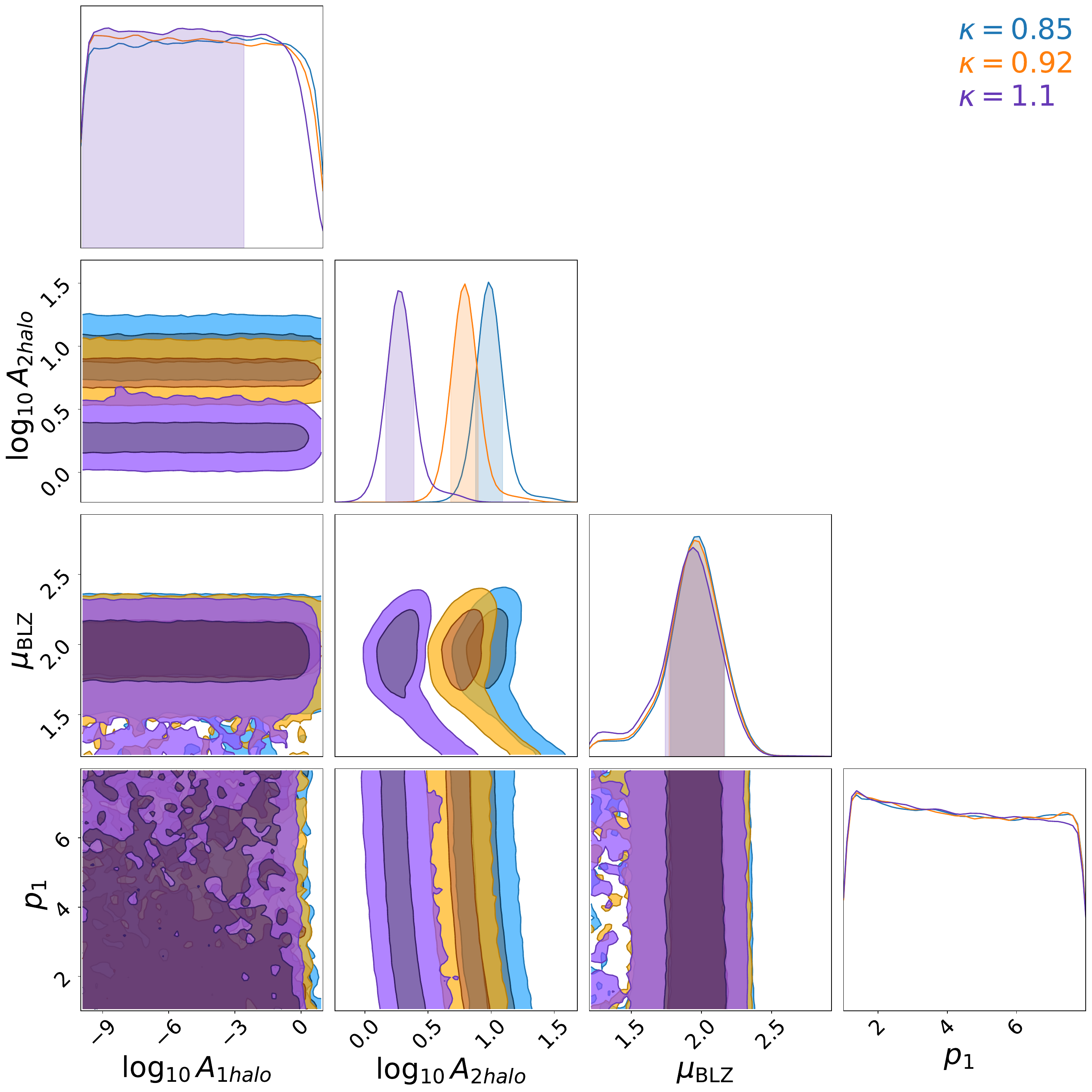}
    \caption{Assessing the impact of the index $\kappa_1$ of the luminosity function on BLZ parameter constraints. The case with $\kappa_1=0.92$ corresponds to the results obtained in Fig.\ \ref{fig:contourplotsphys}.}
    \label{fig:kappacontours}
\end{figure}

As shown in Eq.\ \ref{eq:phies}, the Blazar GLF $\phi_{S}(L_\gamma,z,\Gamma)$ is dependent on the luminosity that (in the faint end probed by this analysis) scales with the luminosity index $\kappa_1$. This parameter is highly degenerate with the normalisation $A$, and therefore we decide not to include an additional parameter in our statistical sampling, assuming the effect of varying $k_1$ would be re-absorbed in the normalization coefficients $A_{\rm 1halo}$ and $A_{\rm 2halo}$. 
In this Appendix, we test such hypothesis by setting $k_1$ to the two extrema of the 68\% interval found in Ref.\ \cite{korsmeier2022flat}, i.e., $\kappa_1=0.92^{+0.18}_{-0.07}$, and repeat the analysis of Section~\ref{subsec:physical_interpretation_galxgam} about cross-correlation of \g-rays with galaxy clustering.
Results are shown in Fig.\ \ref{fig:kappacontours}. 

Considering that the 1-halo component is subdominant and the amplitude is poorly constrained, see Fig.\ \ref{fig:contourplotsphys}, we expect the effect of the normalisation $A$ and of $\kappa_1$ to be mostly encoded by $A_{\rm 2halo}$. 

This is indeed confirmed by Fig.\ \ref{fig:kappacontours}. 
We see that for $\kappa_1=1.1$, $A_{\rm 2halo}$ is reduced compared to the amplitude corresponding to the reference value $\kappa_1=0.92$, owing to the aforementioned degeneracy, i.e., higher $\kappa_1$ means larger $\phi_{S}$ (in the faint end) and thus lower $A_{\rm 2halo}$. In this way, $A_{\rm 2halo}$ becomes closer to unity, even though still somewhat larger.
Decreasing $\kappa_1$ has the opposite effect but to a smaller extent. 
The SNR remains stable ($\sim 5.9$) across all three $\kappa_1$ variations. The parameters $\mu_1$ and $p_1$ show negligible shifts in both constraining power and best-fit values. These results demonstrate that the BLZ model constraints are resilient to reasonable variations in the GLF luminosity index, which are re-absorbed in the normalization parameters introduced in our fits.

\section{Individual Fits for the LP Phenomenological Model}
\begin{figure}
    \centering
    \includegraphics[width=\linewidth]{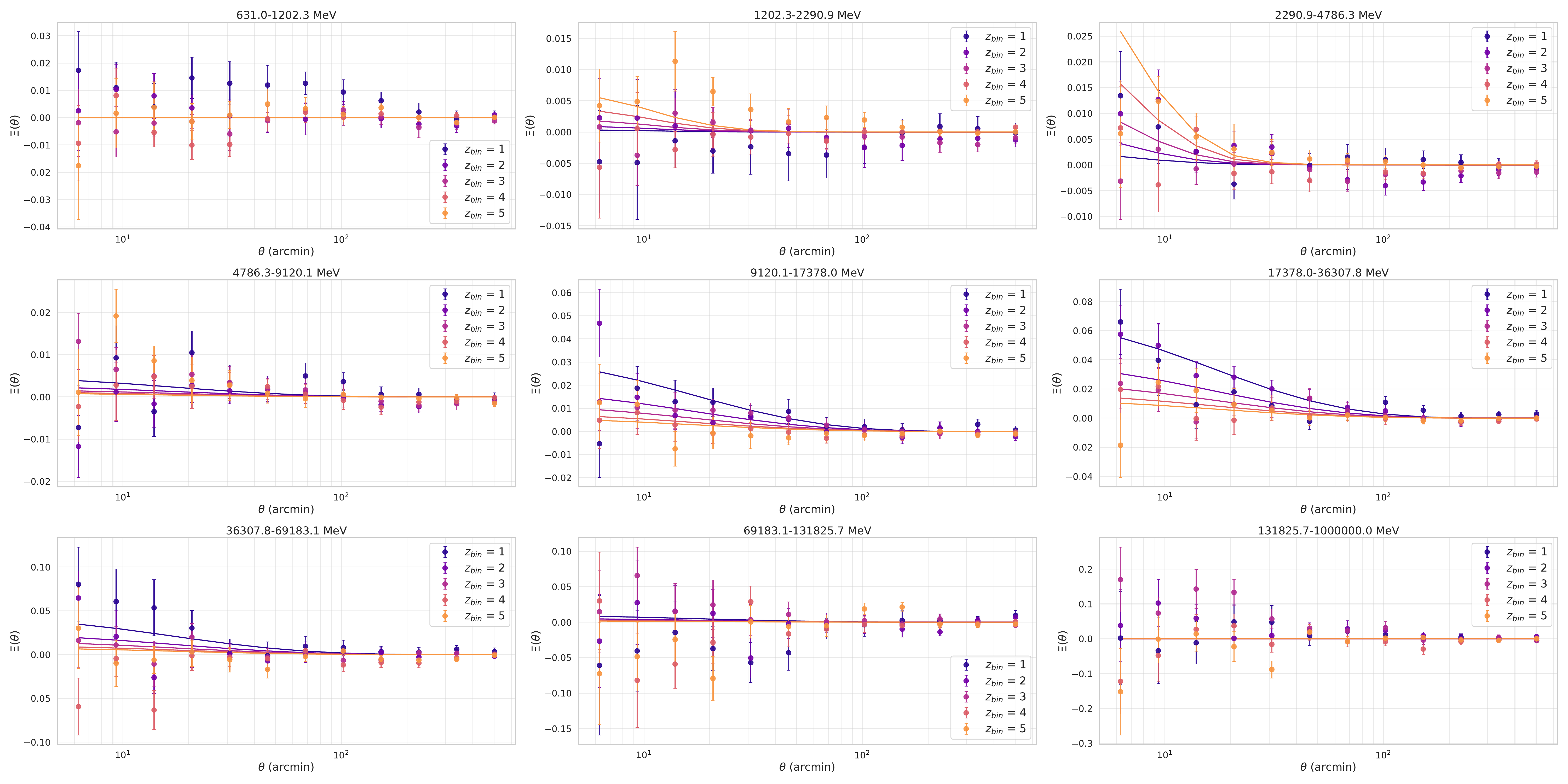}
    \caption{Measurements and LP phenomenological model fits for the cross-correlations between the DES \rmagic\ catalogues and \Fermi\ images, based on the estimator described in Eq.\ \ref{LSestimator_GalxGam}, and reported for all the angular, energy and redshift bins considered in this work.}
    \label{fig:modelallfits}
\end{figure}

In Fig. \ref{fig:modelallfits} we report the breakdown, in terms of energy and redshift bins, of the measured cross-correlation signals together with the best-fit LP phenomenological model. By looking at the different panels, we notice that the LP phenomenological fits are statistically more significant for the intermediate energy bins rather than for very low or very high energy bins, a trend that is also shown for the best-fit LP model in Fig.\ \ref{fig:E_z_thetaplotspheno}.
\end{document}